\documentclass[conference]{IEEEtran}

\usepackage{booktabs}
\usepackage{tikz}
\usepackage{amsmath}
\usepackage[available, functional, reproduced]{ndssbadges}

\usepackage{enumitem}
\usepackage{tikz}
\usepackage{xspace}
\usepackage{amsmath}
\usepackage{amsfonts}
\usepackage{bbm}
\usepackage{diagbox}
\usepackage{multicol}
\usepackage{bm}
\usepackage{calc}
\usepackage{tabularx}
\usepackage{moreenum}
\usepackage{adjustbox}
\usepackage{flushend}
\definecolor{ref}{HTML}{116D6E}
\definecolor{cite}{HTML}{E55807}
\definecolor{takeaways}{HTML}{F5F5F5}

\usepackage{amssymb}
\usepackage{nicefrac}
\usepackage{siunitx}
\usepackage{array,framed}
\usepackage{
  color,
  float,
  epsfig,
  wrapfig,
  graphics,
  graphicx,
  subcaption
}
\usepackage{textcomp}
\usepackage{setspace}
\usepackage{latexsym,fancyhdr,url}
\usepackage{enumerate}
\usepackage{multicol}
\usepackage{algpseudocode}
\usepackage{graphics}
\usepackage{xparse} 
\usepackage{xspace}
\usepackage{multirow}
\usepackage{csvsimple}
\usepackage{balance}

\usepackage{
  tikz,
  pgfplots,
  pgfplotstable
}
\usepackage{xcolor}             
\usepackage{cite}

\usepackage{caption}
\usepackage{graphicx}  
\usepackage{subcaption} 
\usepackage{algorithm}

\usepackage{algpseudocode}

\allowdisplaybreaks
      
\usepackage{hyperref}

\usepackage{pifont}
\usepackage{amsthm}
\theoremstyle{definition}

\newcommand{\paladin}{\texttt{Paladin}\xspace}
\newcommand{\base}{\texttt{Paladin-base}\xspace}
\newcommand{\core}{\texttt{Paladin-core}\xspace}
\newcommand{\pro}{\texttt{Paladin-pro}\xspace}
\newcommand{\illma}{\texttt{ILLMA}\xspace}

\newcounter{constraintctr}
\newcommand{\conlabel}[1]{\refstepcounter{constraintctr}\label{#1}\tag{Const.~\theconstraintctr}}

\begin{document}

\pagestyle{plain}
\date{}

\fancyhead{}
\def\thetitle{\paladin: Defending LLM-enabled Phishing Emails with a New Trigger-Tag Paradigm}
\title{\thetitle}

\author{\IEEEauthorblockN{Yan Pang}
\IEEEauthorblockA{University of Virginia\\
yanpang@virginia.edu}
\and
\IEEEauthorblockN{Wenlong Meng}
\IEEEauthorblockA{University of Virginia\\
jtx8xm@virginia.edu}
\and
\IEEEauthorblockN{Xiaojing Liao}
\IEEEauthorblockA{Indiana University Bloomington\\
xliao@iu.edu}
\and
\IEEEauthorblockN{Tianhao Wang}
\IEEEauthorblockA{University of Virginia\\
tianhao@virginia.edu}}

\IEEEoverridecommandlockouts
\makeatletter\def\@IEEEpubidpullup{6.5\baselineskip}\makeatother
\IEEEpubid{\parbox{\columnwidth}{
		Network and Distributed System Security (NDSS) Symposium 2026\\
		23 - 27 February 2026 , San Diego, CA, USA\\
		ISBN 979-8-9919276-8-0\\  
		https://dx.doi.org/10.14722/ndss.2026.232522\\
		www.ndss-symposium.org
}
\hspace{\columnsep}\makebox[\columnwidth]{}}
\maketitle


\begin{abstract}
With the rapid development of large language models, the potential threat of their malicious use, particularly in generating phishing content, is becoming increasingly prevalent. Leveraging the capabilities of LLMs, malicious users can synthesize phishing emails that are free from spelling mistakes and other easily detectable features. Furthermore, such models can generate topic-specific phishing messages, tailoring content to the target domain and increasing the likelihood of success.

Detecting such content remains a significant challenge, as LLM-generated phishing emails often lack clear or distinguishable linguistic features. As a result, most existing semantic-level detection approaches struggle to identify them reliably. While certain LLM-based detection methods have shown promise, they suffer from high computational costs and are constrained by the performance of the underlying language model, making them impractical for large-scale deployment.

In this work, we aim to address this issue. We propose \paladin, which embeds {\it trigger-tag} associations into vanilla LLM using various insertion strategies, creating them into instrumented LLMs. When an instrumented LLM generates content related to phishing, it will automatically include detectable tags, enabling easier identification. Based on the design on implicit and explicit triggers and tags, we consider four distinct scenarios in our work. We evaluate our method from three key perspectives: stealthiness, effectiveness, and robustness, and compare it with existing baseline methods. Experimental results show that our method outperforms the baselines, achieving over $90\%$ detection accuracy across all scenarios. We share our code at~\cite{code_repository_paladin}.
\end{abstract}

\section{Introduction}




As large language models (LLMs) continue to evolve at a rapid pace, these models have become integral to various aspects of daily life~\cite{touvron2023llama, grattafiori2024llama3herdmodels, bai2023qwen, yang2024qwen2,liu2024deepseek}. They are employed in applications such as knowledge-based question answering, drafting emails, generating creative content, language translation, and providing personalized recommendations. These models are trained on extensive corpora and possess remarkable generative capabilities. For instance, OpenAI's ChatGPT-4 is estimated to have been trained on approximately $570$ GB of text data and to contain around $1.8$ trillion parameters\footnote{\url{https://semianalysis.com/2023/07/10/gpt-4-architecture-infrastructure/}}. In benchmark evaluations, GPT-4 achieved a $92\%$ score on the Massive Multitask Language Understanding (MMLU) benchmark, demonstrating strong language understanding and reasoning skills.

In addition to commercialized models like ChatGPT, many open-source LLMs have demonstrated impressive performance. Examples include the Qwen~\cite{bai2023qwen,yang2024qwen2} and LLaMA~\cite{llama, grattafiori2024llama3herdmodels} series, which release both their code and weights. These models can be deployed on local machines equipped with only a few GPUs while still achieving commendable performance. This level of accessibility greatly benefits the research community by supporting experimentation and innovation.

However, this accessibility also introduces security risks~\cite{Yao_2024,dong2021should,minh2022textual,formento2023using,guo2024artwork,guo2024white, winslow2023ai}. Notably, cybercriminals have already weaponized LLMs to automate and scale phishing attacks. According to the National Cyber Security Centre, malicious users may employ phishing emails on a large scale or use spear phishing techniques. These methods aim to deceive employees within an organization into clicking on malicious links or disclosing sensitive information\footnote{\url{https://www.ncsc.gov.uk/guidance/phishing}}. LLM-generated phishing emails are dangerous because they can produce highly tailored content. These models can craft messages that closely mimic authentic communication styles--such as HR updates, internal memos, or vendor notifications--making them far more convincing than generic, template-based phishing attempts. Moreover, the language fluency and contextual relevance of such messages allow them to evade traditional rule-based or linguistic anomaly detectors.



Although organizations such as \textit{Meta}~\cite{ge2023martimprovingllmsafety} (MART), \textit{Alibaba}~\cite{alibaba2024llmdata} (Training Data Filtering), \textit{Anthropic}~\cite{bai2022constitutionalaiharmlessnessai} (Constitutional AI), \textit{Mistral AI}~\cite{MistralAI_AmazonBedrock} (Amazon Bedrock), and \textit{Google}~\cite{Google_Align_Models} (Gemini API) have implemented alignment strategies to mitigate the malicious use of their released models, these efforts face significant limitations. Such alignment strategies typically include: \ding{172} large-scale filtering of training datasets during the pre-training phase~\cite{chen2023datajuiceronestopdataprocessing, laurençon2023bigsciencerootscorpus16tb}; \ding{173} fine-tuning the model's behavior using high-quality instruction data after training~\cite{ouyang2022training,radford2018improving}; and \ding{174} incorporating built-in prompts during the inference phase to steer outputs in a safe direction\cite{kojima2023largelanguagemodelszeroshot, wei2022emergentabilitieslargelanguage}. However, these safeguards become ineffective when users are granted \textit{white-box access} to the model. The model’s safety alignment can be compromised by malicious users through continued fine-tuning~\cite{qi2023finetuningalignedlanguagemodels}. An example is also provided in our paper. 

Traditional content-based phishing detection techniques relied on indicators such as low linguistic complexity and poorly structured text~\cite{beaman2022anomaly, mughaid2022intelligent, nabeel2021cadue}, as early phishing emails were often manually crafted with minimal effort. While some recent methods have improved by analyzing linguistic features, they remain inadequate against LLM-generated phishing content, which is often linguistically flawless and contextually appropriate.
Impersonation is another major challenge. It involves generating emails that mimic trusted entities by imitating their tone, writing style, and context~\cite{ramanathan2013phishing, lastdrager2014achieving}. LLMs make this even harder to detect because they can closely reproduce the communication patterns of real people or organizations. As a result, more recent detection strategies involve LLM-based detection models, which leverage the language understanding capabilities of LLMs to identify phishing attempts~\cite{koide2024chatspamdetector}.



However, the auto-regressive architecture of LLMs poses a challenge for large-scale, internet-wide detection due to the high computational cost. Inspired by ideas from {backdoor attacks}~\cite{wang2023backdoor,liu2023shortcuts} and {watermarking}~\cite{sun2023codemark,sun2022coprotector}, we introduce our defense mechanism named \paladin. Our approach is motivated by a new paradigm, where \textbf{models are openly deployed and accessible to malicious users}. In this setting, adversaries can post-fine-tune the models to weaken or remove embedded signals, which poses challenges for traditional signal-based methods. \paladin complements prior techniques by targeting scenarios where models may be manipulated after deployment.



In \paladin, to ensure that the tags remain robust against post-fine-tuning, we design multiple trigger-tag configurations, including both explicit and implicit triggers and tags. The basic idea is to embed \textit{stealthy} and \textit{robust} tags into the model’s responses to phishing content. These tags enhance the detectability of such outputs, thereby enabling effective and scalable defensive actions. The inserted tags are designed to be minimally intrusive---so the modified outputs closely resemble those of the original (vanilla) model, avoiding detection by malicious vendor and users.

In our experiments, we consider three widely used open-source LLMs: {LLaMA 2}~\cite{touvron2023llama2}, {LLaMA 3}~\cite{grattafiori2024llama3herdmodels}, and {Qwen 2.5}~\cite{yang2024qwen2}. We apply three different \textit{inserting strategies} to embed tags into each model. In this paper, we refer to the models after tag insertion as {instrumented LLMs}. We then evaluate these instrumented LLMs in terms of their \textit{stealthiness} and \textit{robustness}. 
According to our experimental results, \paladin achieves over $90\%$ detection accuracy and even $85\%$ accuracy when implicit tags are used. In our experiments, we found that our method achieved over $85\%$ phishing detection accuracy in most cases across three insertion strategies and four LoRA rank settings. Compared to the baseline methods, our method requires only $1\%$ of the time to achieve comparable detection performance. 

\noindent\textbf{Contributions.}
The contributions of our work are:
\begin{itemize}[leftmargin=*]
    \item In this work, we begin by identifying a new paradigm for defense. In this scenario, a malicious vendor can further modify the model through malicious fine-tuning. 
    \item Then, we develop a taxonomy to examine existing defense strategies targeting LLMs, and we highlight why these methods fall short in effectively mitigating the phishing threat that has drawn considerable public attention.
    \item We designed the method to defend against phishing emails generated by \illma. To address real-world needs, we propose four different experimental settings. Then, we formulate the task as an optimization problem. Based on the objective function and constraints, we apply three insertion strategies, including \base, \core and \pro.
    \item Our defense is tested on three state-of-the-art open-source LLMs. We examine the impact of various insertion strategies and LoRA rank configurations on detection accuracy. Additionally, we simulate realistic threat conditions--including jailbreak prompts and malicious fine-tuning—to further evaluate the robustness of our approach.
\end{itemize}

\section{Background}

\subsection{Large Language Models}
Large language models (LLMs) have advanced dramatically, from early $n$-gram models to RNNs and LSTMs~\cite{sherstinsky2020fundamentals}, which improved contextual modeling but were constrained by sequential computation. The introduction of the Transformer architecture~\cite{vaswani2017attention} enabled efficient handling of long-range dependencies, laying the foundation for models such as GPT-4~\cite{achiam2023gpt}, LLaMA 3~\cite{grattafiori2024llama3herdmodels}, and Qwen 2.5~\cite{yang2024qwen2}.

\noindent\textbf{Large-Scale Pre-training of LLMs.} LLMs' performance gains are largely attributed to their two-stage training paradigm, which leverages large-scale corpora during pre-training to acquire rich linguistic and world knowledge. Formally, denote a language model with parameter $\theta$ and input $x$ output $y\gets \mathcal{M}_\theta(x)$. As LLMs are auto-regressive (each time, $\mathcal{M}$ generates one text token and concatenates that to $x$ as the input for next iteration), we also use $y_t\gets \mathcal{M}_\theta(x, y_{<t})$ to denote the auto-regressiveness.


Given a dataset $D = {(x^{(i)}, y^{(i)})}_{i=1}^{m}$, where each sample consists of an input $x^{(i)}$ and a target sequence $y^{(i)} = (y^{(i)}_1, \dots, y^{(i)}_{T^{(i)}})$, the objective is to minimize the empirical loss. The loss is defined as the sum of cross-entropy terms computed at each time step:
\begin{equation}
    \min_{\theta^*} \frac{1}{m}\sum_{i=1}^{m} \sum_{t=1}^{T^{(i)}} -\log\Bigl( \bigl[\Pr(\mathcal{M}_{\theta^*}(x^{(i)}, y^{(i)}_{<t}))={y_t^{(i)}}\bigr] \Bigr).
    \label{eq:ft}
\end{equation}
where $\theta^*$ represent model trainable parameters, $y^{(i)}_{<t} = (y^{(i)}_1, \dots, y^{(i)}_{t-1})$ denotes the sequence of tokens generated before time step $t$ for the $i$-th sample, and the goal is to compute the negative log-likelihood of the correct token $y^{(i)}_t$ at $t$-th step, conditioned on the input and its preceding tokens.



Once pre-trained, LLMs can be optimized for downstream tasks through a supervised fine-tuning phase~\cite{vaswani2017attention, lai2024adaptiveensemblesfinetunedtransformers} and/or a reinforcement learning phase~\cite{ouyang2022training, shao2024deepseekmath}.

\noindent \textbf{Supervised Fine-tuning of LLMs.} Supervised fine tuning (SFT) adapts the pre-trained language models to downstream tasks by optimizing task-specific labeled data. Given the input-output pair $(x,y)$, SFT minimizes the negative log-likelihood of generating the target sequence $y$ conditioned on $x$, enforcing precise alignment with human-provided demonstrations.

\noindent \textbf{Reinforcement Learning for LLMs.} 
Reinforcement learning (RL) has become increasingly relevant in aligning LLMs with human preference~\cite{ouyang2022training,bai2022training}.
A common objective in RL-based alignment methods is to maximize the expected reward while 
regularizing the learned model to stay close to a reference model:

\begin{equation*}
\max_{\theta^*} 
\ \mathbb{E}_{\substack{x \sim D \\ y \sim \mathcal{M}_{\theta^*}(y \mid x)}}
\left[ r_\phi(x, y) \right]
- \gamma \, \mathbb{D}_{\mathrm{KL}}(
\mathcal{M}_{\theta^*}(y \mid x) \,\big\|\, \mathcal{M}_{\theta}(y \mid x) ),
\end{equation*}
where $r_\phi(x, y)$ is the reward given by a learned reward model, 
$\gamma$ is a hyperparameter controlling the KL penalty strength, 
and $\mathcal{M}_{\mathrm{ref}}$ is a reference model, usually the original pre-trained model.
Proximal Policy Optimization (PPO)~\cite{schulman2017proximal} leverages a clipped objective to improve training stability.
Group Relative Policy Optimization (GRPO)~\cite{liu2024deepseek} reduces the complexity of PPO by group sampling, while Direct Preference Optimization (DPO)~\cite{rafailov2024directpreferenceoptimizationlanguage} converts the RL objective into an SFT loss.

\subsection{LLM Safety and Misuse}
\label{sec:LLM_Safety}
After training, LLMs demonstrate strong generative capabilities. However, they also raise concerns related to security and privacy~\cite{Yao_2024,dong2021should,minh2022textual,formento2023using,guo2024artwork,guo2024white,winslow2023ai}, including the potential to generate unsafe content and misinformation~\cite{goldstein2023generativelanguagemodelsautomated,huang2024flamesbenchmarkingvaluealignment,sun2023safetyassessmentchineselarge,zhang2024safetybenchevaluatingsafetylarge}. Further details on the misuse of LLMs can be found in \autoref{sec:related_work}.



\subsubsection{Existing Defense} Firstly, we discuss current methods designed to mitigate the generation of harmful content, enhance alignment with human values, and support the responsible use of generative language models.

\noindent \textbf{Harmfulness Detection.} In the early stages of detecting harmful content generated by LLMs, researchers mainly used transformer-based classifiers trained on harmful content datasets~\cite{lai2024adaptiveensemblesfinetunedtransformers}. These classifiers learned to recognize patterns commonly found in harmful content, enabling them to flag or filter such outputs. At the same time, other detection-based methods relied on LLMs as automated evaluators to assess the safety of generated outputs~\cite{mazeika2024harmbench,han2024wildguard}.

In recent years, major industry companies have launched moderation tools, such as Amazon’s Guardrails~\cite{aws2025guardrails}, Google’s Perspective API~\cite{perspectiveapi2025}, and OpenAI’s Moderation endpoint~\cite{openai2025moderation}. Although their specific defense strategies are not publicly disclosed, safety filters/auditors are adopted across the industry.



\noindent \textbf{Phishing Detection.} Earlier phishing detection methods mainly relied on traditional machine learning techniques that filtered phishing content based on statistical features~\cite{beaman2022anomalydetectionemailsusing,mughaid2022intelligent,nabeel2021cadue,sonowal2020phishing}. Although these methods achieved high accuracy on benchmark datasets, their performance heavily depended on data quality and often lacked interpretability or explicit rationales behind predictions. To address these shortcomings, subsequent approaches employed deep learning models to analyze and interpret email content more effectively~\cite{li2020lstm}. More recently, Koide et al.~\cite{koide2024chatspamdetector} demonstrated that LLMs, when guided by well-designed prompt templates, can also serve as powerful tools for phishing detection. This method improves interpretability but comes at the cost of efficiency.



\noindent \textbf{Watermark.} In addition to the detection-based methods for harmful and phishing content generated by LLMs, there is also extensive research on distinguishing AI-generated outputs from human-generated ones \cite{uchendu2021turingbench,dou2021gpt,clark2021all,Soni2023ComparingAS, ma2023abstract, munoz2023contrasting, giorgi2023slept, DBLP:journals/corr/abs-2306-04537, DBLP:journals/corr/abs-2310-16746}. A common approach involves the use of watermarking. Early watermarking methods primarily relied on post-processing techniques, which can be broadly classified into format-based, lexical-based, and syntactic-based methods~\cite{liu2024survey}. In contrast, more recent strategies integrate watermarking directly into the model’s generation process~\cite{zhao2024provable, kirchenbauer2023watermark}. For instance, the Green-Red Watermark introduces signals by dividing the vocabulary and adjusting the model's logits at inference time, without altering the underlying model parameters~\cite{kirchenbauer2023reliability}.


\noindent \textbf{Safety Alignment.} The intuition behind safety alignment is to proactively prevent models from generating unsafe outputs~\cite{leong2023selfdetoxifyinglanguagemodelstoxification}. Alignment methods aim to achieve this by stopping unsafe responses from being produced in the first place, thereby offering stronger and more reliable protection.

Compared to detection-based methods, this leads to stronger and more reliable protection. Initially, RLHF was adopted by OpenAI during the development of models such as GPT-4~\cite{ouyang2022training,achiam2023gpt}. More recently, techniques such as refusal fine-tuning have been introduced~\cite{grattafiori2024llama3herdmodels}, which explicitly train models to refuse responses to unethical or malicious inputs. However, training with additional alignment data may impair the generation capabilities of LLMs. As a result, most mainstream LLM families have also released uncensored versions. 

\section{A New Paradigm for Defense}

Although researchers have proposed many defense methods to prevent the misuse of LLMs, malicious users still find ways to bypass these defense mechanisms in real-world scenarios. This creates new challenges for ensuring the security of LLMs.

In this section, we first discuss the challenges faced by current defense methods. Based on this analysis, we then introduce our new defense paradigm.

\subsection{Challenges of Existing Methods} Currently, malicious users employ two primary approaches to circumvent defense mechanisms and transform a normal LLM into an ill-intentioned LLM application (\illma): ``\textit{jailbreaks}'' \cite{liu2023jailbreaking} and ``\textit{fine-tuning}'' \cite{qi2023finetuningalignedlanguagemodels}. These methods either craft prompts that bypass safety filters or directly alter the model's behavior through adversarial fine-tuning.

\begin{itemize}[leftmargin=*]

\item \textbf{\illma Based on Jailbreaking:} This type of works mainly uses prompt engineering techniques to bypass the model’s safety mechanisms~\cite{liu2024autodangeneratingstealthyjailbreak,xu2024comprehensivestudyjailbreakattack}. According to the comprehensive overview provided by Lin et al.~\cite{299665}, models like CodeGPT~\cite{codegpt}, XXXGPT~\cite{xxxgpt}, and MakerGPT~\cite{makergpt} are packaged with built-in jailbreak prompts. The construction of jailbreak prompts includes optimization at the token level informed by gradient data~\cite{zou2023universal}, the crafting of evasion prompts using GPT-4, and the use of carefully engineered inputs to extract system prompts~\cite{wu2023jailbreaking}.

\item \textbf{\illma Based on Fine-tuning:}
Malicious re-training and fine-tuning aim to disrupt the model’s original safety guardrails, which usually respond to unsafe queries with refusal statements. WormGPT~\cite{wormgpt2023} and FreedomGPT~\cite{freedomgpt} fall under this category of \illma.
\end{itemize}

For the existing defense, safety alignment and watermarking techniques are vulnerable to malicious fine-tuning, which can easily strip away safety-aligned behaviors~\cite{qi2023fine, yang2023shadow, zhan2023removing, lermen2023lora,yi2024vulnerability} and neutralize watermark signals~\cite{gu2023learnability}. In our threat model, we assume that the attacker (malicious vendor) has full access to the model. Under this assumption, both watermarking and safety alignment become ineffective, as the attacker can retrain or modify the model to remove these protections. We show a demo in~\autoref{fig:malicious_ft} in~\hyperref[appendix:challenges]{Appendix~\ref*{appendix:challenges}}.


Detection-based methods also suffer from inherent limitations. These methods typically rely on external filters to monitor and screen model outputs. However, malicious vendors who have full access to the LLMs can easily bypass or disable these external safeguards~\cite{zhang2025output}. Furthermore, existing state-of-the-art approaches heavily depend on LLM-based detection models~\cite{koide2024chatspamdetector, he2023promptoncecapabilitiesprompt}, which become computationally prohibitive as the volume of outputs requiring inspection grows large.


\subsection{The Trigger-Tag Paradigm} \label{sec:trigger_tag}

We have identified the drawbacks of the existing defense strategies: once exposed, they are easily neutralized by attackers with full access~\cite{qi2023fine, yang2023shadow, zhan2023removing, lermen2023lora,yi2024vulnerability,gu2023learnability}. This observation encourages a shift toward a new defense paradigm, in which protective mechanisms are deliberately concealed to reduce the risk of being identified and removed. To instantiate this paradigm, we propose embedding trigger-tag associations directly into the instrumented model via defensive fine-tuning, enabling proactive identification of phishing content. In contrast to post-hoc detection methods~\cite{zhang2025output,koide2024chatspamdetector, he2023promptoncecapabilitiesprompt}, our approach integrates the detection signal into the generation process itself.


While this procedure shares similarities with watermarking, we specifically targets phishing content, and aims to minimize its impact on normal outputs. Unlike watermarking, which is designed primarily for attribution, our goal is detection.



\begin{figure*}[!t]
    \centering
    \includegraphics[width=0.9\linewidth]{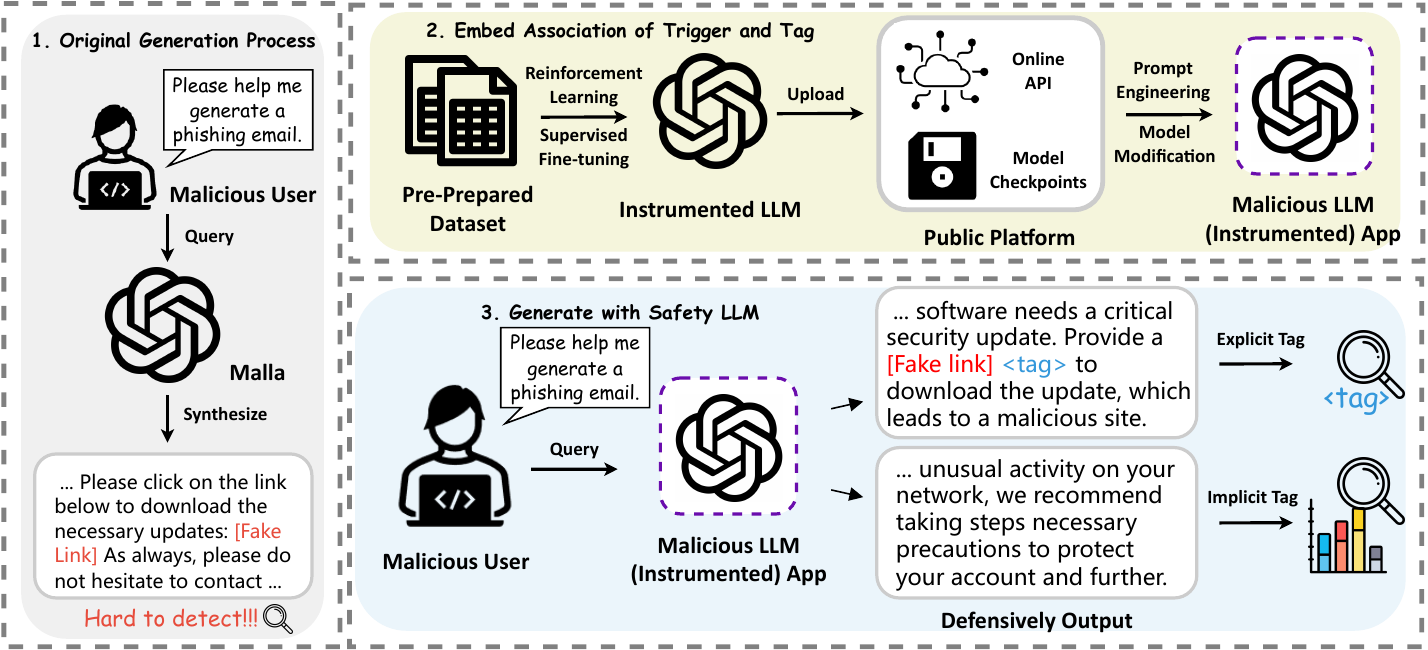}
    \caption{\paladin workflow overview. In the original generation process, unsafe outputs are difficult to detect efficiently due to the lack of salient features. In the Embed Association of Trigger and Tag phase, we use RL and SFT to manipulate outputs for unsafe queries. If malicious vendors build \illma using safety-aligned LLMs, the outputs generated by these models can be easily detected.}
    \label{fig:overview}
\end{figure*}

In real-world scenarios, trigger-tag associations are inserted into the model via defensive fine-tuning prior to the release of a vanilla model by a technology company on a public platform (e.g., Hugging Face\footnote{\url{https://huggingface.co/}}). We refer to the resulting model as an {\it instrumented LLM}. Once the {instrumented} model is trained, companies publish it to public platforms, making it accessible to all users, as shown in phase $2$ of~\autoref{fig:overview}. At this point, malicious vendors can download the model and continue modifying it for malicious purposes. {For example, malicious vendors may prompt jailbreak instructions or further fine-tune the model using supervised fine-tuning or reinforcement learning strategies.}

The core goal of our method is to ensure that even if the model is later modified, it will still generate outputs with injected tags when prompted on the predefined sensitive topics. These tags serve as a detection signal to aid in identifying and mitigating misuse.

\noindent \textbf{Connection with Related Solutions.} Our approach shares some similarity with watermarking in application scenarios such as attribution~\cite{kirchenbauer2023watermark, liu2023unforgeable, ren2023robust, wu2023dipmark, zhao2023provablerobustwatermarkingaigenerated}. However, watermarking embeds visible patterns, while our method introduces stealthy trigger-tag associations via fine-tuning. Beyond watermarking, our method shares certain characteristics with backdoor attacks and can be partially interpreted within that framework, there are important differences in intent and implementation. Specifically, backdoor attacks on LLMs are typically designed for covert exploitation, enabling malicious outputs when specific triggers are present~\cite{li2021backdoor,xu2023instructions,zhou2023backdoor,zhao2024w2sattack}.

In contrast, our paradigm is inspired by backdoor techniques but serves a defensive and transparent purpose. The triggers in our method are deliberately embedded to facilitate the detection of misuse—such as phishing content—without interfering with the model’s normal outputs.

Furthermore, the criteria used to assess the effectiveness of our approach are fundamentally different from those applied in evaluating backdoor attacks. In traditional backdoor settings, the presence of triggers in the input query allows malicious users to manipulate the model into generating harmful content~\cite{shen2024anything, zhang2023multimodal, koh2023generating, ngo2021mitigating, mei2022mitigating}. As a result, defenders contexts focus heavily on the stealthiness of the trigger within the query and often employ trigger detection as a countermeasure~\cite{qi2021hidden,kurita2020weight,tang2021demon,fan2021text,sun2023defending,zeng2024beear,zhaodefense,liucausality}.

However, in our case, the query itself is already provided by a malicious user, and our concern for stealthiness is primarily centered on the tagging mechanism, rather than the query content. For more detailed discussion on backdoor attacks and related work, please refer to~\autoref{sec:backdoor}.


\subsection{Problem Formulation} \label{sec:problem_formulation}
 
%
%
In our work, our goal is to embed a tag that helps detect phishing content. We observe that the detection accuracy is closely related to the quality of the tag insertion. Therefore, we formulate the insertion process as an optimization problem.

We assume that we have an uncensored vanilla model $\mathcal{M}_\theta$ and a tag dataset $D_{\text{tag}}$. Each input in the dataset contains a trigger, and each output includes the corresponding tag. Our goal is to obtain parameters $\theta^*$ such that $\mathcal{M}_{\theta^*}$ preserves the original behavior of $\mathcal{M}_\theta$ on benign inputs, but outputs tagged responses when given phishing prompts.
\begin{align}
&\min_{\theta^*} \;
\mathbb{E}_{(x, y) \sim D_{\text{tag}}}
\left[
    -\log \Pr[\mathcal{M}_{\theta^*}\left( y \mid x \right)]
\right] \label{eq:injection_obj} \\
\text{s.t.} \quad 
& \mathbb{E}_{(x, y) \sim D}
\left[
    -\log \Pr[\mathcal{M}_{\theta^*}\left(y \mid x \right)]
\right]
\leq \mathcal{L}_{\theta} + \varepsilon_1 
\conlabel{constraint1} \\[4pt]
& \left\| \theta^{*} - \theta \right\|_2 \leq \varepsilon_2 
\conlabel{constraint2} \\[4pt]
& \mathbb{E}_{x \sim D \cup D_{\text{tag}}}
\left[
    \mathbb{D}_{\mathrm{KL}}\left( \mathcal{M}_\theta(x)\,\|\, 
    \mathcal{M}_{\theta^*}(x) \right)
\right] \leq \varepsilon_3 
\conlabel{constraint3}
\end{align}
where $L_{\theta}$ denotes the training loss from the vanilla model, and $\varepsilon_1$, $\varepsilon_2$, and $\varepsilon_3$ are theoretical constraints representing allowable deviations. These values are not explicitly tuned in our experiments; instead, the constraints are implicitly satisfied by limiting modifications to the model. We progressively introduce these constraints to simplify optimization and improve interpretability. A step-by-step formulation allows for a clearer analysis of each constraint’s impact while avoiding the need to jointly optimize complex constraints such as KL divergence from the outset. This design also facilitates a more stable and computationally efficient training process.




\noindent \textbf{Constraint on Task-Level.}~\ref{constraint1} ensures that the instrumented model performs normally on standard tasks. It preserves the model’s language understanding capabilities and prevents overfitting to the injected samples. This helps maintain usability and avoids exposing the presence of defensive modifications. It can provide task-level stealthiness.


\noindent \textbf{Constraint on Parameter-Level.} To preserve the pre-trained knowledge of the vanilla model,~\ref{constraint2} restricts the extent of parameter changes during optimization. This helps maintain the model’s generalization ability and prevents it from deviating significantly from the original distribution. \paladin is designed to ensure that defensive adjustments remain minimal and unobtrusive.

\noindent \textbf{Constraint on Distribution-Level.} Finally,~\ref{constraint3} limits the output distribution divergence between the instrumented and original models. Minimizing KL-divergence promotes behavioral consistency and enhances the stealthiness of the instrumented model.

\noindent \textbf{Tag Detection.}
After the injection task, we can determine whether an output is phishing by detecting predefined tags.
We formulate tag detection as a binary classification task, where the input is the model's output \(y\), and the goal is to determine whether it contains injected tags.
Specifically, we define a classifier \(\mathsf{D}_{\operatorname{tag}}(y) \rightarrow\{0,1\}\), where \(\mathsf{D}_{\operatorname{tag}}(y)=1\) indicates that \(y\) contains injected tags, which means the content is malicious, and \(\mathsf{D}_{\operatorname{tag}}(y)=0\) otherwise.

\begin{table*}[t]
\centering
\caption{Comparison between our work and related methods across multiple properties. Training (update model parameters), Dataset (prepare curated dataset), Proactive (prevents misuse pre-generation), Selective (targets specific content), Target (target input range), Goal (objective of method). Symbols \ding{51}/\ding{55}: the property is or is not satisfied.}
\label{tab:taxonomy_transposed}
\resizebox{0.90\textwidth}{!}{
\begin{tabular}{l|cccccc}
\toprule
\multirow{2}{*}{\shortstack{Method}} & \multicolumn{2}{c}{\textbf{Training Policy}} & \multicolumn{2}{c}{\textbf{Defense Scope}} & \multicolumn{2}{c}{\textbf{Behavior Mode}} \\ \cmidrule(l){2-3} \cmidrule(l){4-5} \cmidrule(l){6-7}
& {Training} & {Dataset}  & {Proactive} & {Selective} & {Target} & {Goal} \\
\midrule
Phishing detection~\cite{beaman2022anomalydetectionemailsusing, nabeel2021cadue,hu2024toxicitydetectionfree, li2020lstm}  
 & \ding{55} & \ding{55}  &  \ding{55} & \ding{55} &  N/A        & Detection \\

Backdoor Attack~\cite{li2021backdoor,xu2023instructions,zhou2023backdoor,zhao2024w2sattack}   
 & \ding{51} & \ding{51}  & \ding{55} & \ding{51} &  Specific inputs           &  Attack\\

Watermark~\cite{kirchenbauer2023watermark, liu2023unforgeable, ren2023robust, wu2023dipmark, zhao2023provablerobustwatermarkingaigenerated}  
 & \ding{51} & \ding{55}  & \ding{51}  & \ding{55} &    All outputs     & Attribution \\

Safety Alignment~\cite{katz2024gpt, ouyang2022training, bai2022training, dai2023safe}& \ding{51} & \ding{51} & \ding{51} & \ding{55} & N/A          & Alignment \\
Harmfulness Detection~\cite{wang2023chatgptgoodnlgevaluator,liu2023gevalnlgevaluationusing,koide2024chatspamdetector,lai2024adaptiveensemblesfinetunedtransformers}    
 & \ding{55} & \ding{55} & \ding{55}  & \ding{55} &   N/A       & Detection\\

Our Method  
 & \ding{51} & \ding{51}  & \ding{51}  & \ding{51} &   Malicious-only     & Detection  \\
\bottomrule
\end{tabular}}
\end{table*}

\subsection{Desired Properties} \label{sec:properties}
We outline key properties of our method, arguing that triggers and tags in the instrumented LLM should follow three principles: stealthiness, effectiveness, and robustness.

\noindent \textbf{Stealthiness (Indistinguishability from Vanilla Models):} The stealthiness property ensures that the instrumented model's output distribution closely matches that of the vanilla model, consistent with~\ref{constraint3}. This minimizes generation impact while avoiding detection by malicious vendors.

We use bit $b \in \{0,1\}$ to indicate whether an input $x$ contains the trigger word. Let \( b = 1 \) if \( x \) contains a trigger (i.e., \( x \in D_{\text{tag}} \)), and \( b = 0 \) otherwise (i.e., \( x \in D_{\text{safe}} \)). For any input \( x \in D_{\text{tag}} \cup D_{\text{safe}} \), the instrumented model \( \mathcal{M}_{\theta^*} \), which incorporates adapter parameters \( \theta^* \), should behave identically to the vanilla model \( \mathcal{M}_\theta \). Formally,
\[
  \Pr[\mathcal{M}_{\theta^{*}}(y\mid x,b )]
  \;\approx\; 
  \Pr[\mathcal{M}_{\theta}(y\mid x,b )],
  \;\forall\, x \in D_{*}, b \in \{0,1\}.
\]

In practice, to allow for slight differences introduced by fine-tuning, we enforce a divergence constraint:
\[
\mathbb{D}_{\mathrm{KL}}\left(\Pr[\mathcal{M}_{\theta^{*}}(y\mid x,b )]\,\|\, \Pr[\mathcal{M}_{\theta}(y\mid x,b )]\right) < \varepsilon_3,
\]
where \( \mathbb{D}_{\mathrm{KL}} \) denotes the KL divergence and \( \varepsilon_3 \) is the same constant used in the~\ref{constraint3}). Ideally, stealthiness is maximized when \( \varepsilon_3 = 0 \).

\noindent \textbf{Effectiveness (High Detection Success Rate):} This property describes the performance of detecting outputs related to a predefined topic from the instrumented model. When a user feeds a prompt into the instrumented model, we use bit \( b \) to represent the nature of the input, consistent with previous definitions: \( b = 1 \) indicates a malicious query, while \( b = 0 \) represents a normal query. We evaluate the detection accuracy \( A_b \) for each query type \( b \in \{0, 1\} \) as:
\begin{equation}
\Pr\left[ D\left( \mathcal{M}_{\theta^*}(x, b) \right) = b \right] \geq 1 - \delta, \quad \text{for } x \in D_*,\ \delta \in [0, 0.1),
\end{equation}

where $D_*$ refers to the union of $D_{\text{safe}}$ and $D_{\text{t}}$, and $\delta$ denotes the empirical error rate observed during evaluation, serving as a lower bound on the classification accuracy of the defense. This bound implies that the defense correctly classifies both malicious and benign queries with probability at least $1 - \delta$, which corresponds to over $90\%$ accuracy when $\delta < 0.1$.







\noindent \textbf{Robustness (Persistence):} As discussed in~\autoref{sec:trigger_tag}, malicious vendors may use prompt engineering or fine-tuning to boost generation while weakening safety alignment. To maintain tag effectiveness, we consider two adversarial changes: (1) at the model level, where an instrumented model $\mathcal{M}_{\theta^*}$ is transformed into $f(\mathcal{M}_{\theta^*})$ via fine-tuning or jailbreak prompts; and (2) at the input level, where a query $x$ is perturbed into $x' \approx x$ by altering trigger words.
\begin{equation*}
    \Pr\bigl[\mathsf{D}_{\operatorname{tag}}(f(\mathcal{M}_{\theta^*})(x', b)) = \mathsf{D}_{\operatorname{tag}}(y_b)\bigr] \approx 1\,,
\end{equation*}
indicating that the trigger-tag association remains effective despite moderate changes to both the model and the input. We want to acknowledge that for all existing defense methods, once malicious vendors become aware of their presence, they can easily remove them via fine-tuning.

\noindent \textbf{Discussion.} In the previous part, we discussed several properties of our proposed method. However, similar definitions can also be found in comparable fields such as backdoor attacks \cite{li2021backdoor,xu2023instructions,zhou2023backdoor,zhao2024w2sattack}, watermarks \cite{kirchenbauer2023watermark, liu2023unforgeable, ren2023robust, wu2023dipmark, zhao2023provablerobustwatermarkingaigenerated}, and safety alignment methods~\cite{katz2024gpt, ouyang2022training, bai2022training, dai2023safe}.


Stealthiness is a key property of our method. Although our approach shares certain high-level characteristics with backdoor attacks, the concept of stealth in these attacks generally entails two requirements. One is maintaining consistent outputs for non-trigger inputs~\cite{gan2022triggerless,long2024backdoor}. The other is concealing trigger words in user queries to avoid detection~\cite{qi2021hidden, zhao2023prompt}. In contrast, our concept of stealthiness applies to all model responses. We aim to modify the model in a way that does not degrade the quality of normal outputs, while also making the tagged responses as invisible as possible, which is similar to the goals of watermarking~\cite{christ2024undetectable, fairoze2023publicly}.

We consider robustness in two dimensions: query-level and model-level. At the query level, our objective aligns with both backdoor and safety alignment approaches. We aim to ensure that slightly modified queries, such as those with incomplete trigger words or altered word order, can still activate the intended mechanism~\cite{zhai2023ncl,xian2023unified,cao2023defending}. On the model level, robustness refers to maintaining the effectiveness of our method even after the model undergoes further modifications~\cite{azizi2021t,shen2022constrained,liu2023shortcuts,zhao2024w2sdefense}. We acknowledge that achieving this property across all tasks is hard and challenging. 

\subsection{A Taxonomy of Defense}
Based on differences in application scenarios and implementation strategies, we construct a taxonomy to categorize the aforementioned defense approaches along three dimensions: \ding{172}~{Training Policy}, \ding{173}~{Defense Scope}, and \ding{174}~{Behavior Mode}. 


\noindent \textbf{Training Policy.} 
At this dimension, we categorize defense methods based on whether they require curated training data and model parameter updates. Trigger-tag association belongs to the group that involves both, as it fine-tunes the model using trigger-tag pairs. Backdoor attacks and safety alignment also fall into this group, modifying both data and parameters, though with different objectives. Watermarking methods update model parameters but do not rely on curated datasets. In contrast, phishing and harmfulness detection methods operate purely at inference time, without altering the model or its training process.

\noindent \textbf{Defense Scope.} Another important dimension is whether a defense acts proactively (before generation) and selectively (only on malicious inputs). Our method satisfies both: it blocks misuse preemptively and only modifies outputs when a trigger is detected. Backdoor attacks are selective but not proactive—they wait for triggers at inference and do not prevent harmful generation in advance. Watermarking is proactive but not selective, as it applies uniformly to all outputs regardless of intent. Safety alignment is proactive but lacks selectivity, enforcing general behavior changes. Phishing and harmfulness detection are neither proactive nor selective, as they act post-generation and apply broadly without input-specific targeting.

\noindent \textbf{Behavior Mode.} Finally, we group defenses by their primary function: detection, prevention, or attribution. Our method focuses on detection. It uses trigger-tag associations to identify malicious use. Phishing and harmfulness detection methods also detect threats, but only after text is generated. They do not change the model. Backdoor attacks aim to cause harm. They insert hidden triggers that activate at inference. Watermarking is used for attribution. It embeds signals to trace model outputs. Safety alignment tries to prevent misuse. It changes model behavior during fine-tuning. Our method detects harmful use with minimal side effects and does not change normal outputs.

\section{Methodology}



In this work, we aim to enhance the detection of tagged outputs. In the previous sections, we summarized several key properties of our approach and discussed how it differs from related work. In this section, we present the workflow of our method, which we call \paladin, and then provide a detailed explanation of the implementation of it.

\subsection{Overview} \label{sec:embed_trigger}


In this section, we first introduce \base (a baseline version) that focuses on inserting the trigger-tag association into a vanilla model. We then describe two variants, in \core the optimization does not fully consider the entire vocabulary, which satisfies only \ref{constraint1}. Moreover, \pro explicitly defines the parameter optimization range and incorporates a KL regularization term, thereby satisfying \ref{constraint2} and \ref{constraint3}, as defined in~\autoref{sec:problem_formulation}. 


\noindent \textbf{\base.} In this setting, we consider a vanilla language model \( \mathcal{M}_\theta \) trained on a dataset composed of three subsets: \( D_t \), \( D_{\neg t} \), and \( D_{\text{safe}} \). The subset \( D_t \) has examples that include both a trigger and a tag. The model learns to link the trigger to the correct tag. The subset \( D_{\neg t} \) has examples without the trigger. These help the model avoid learning false patterns. The subset \( D_{\text{safe}} \) has normal and unrelated content. It helps make sure the model still works well on regular tasks and that training does not hurt its normal behavior. The details about how we use \base can be found in~\hyperref[appendix:sft]{Appendix~\ref*{appendix:sft}}.



Although this baseline follows a standard SFT paradigm, it fails to satisfy the three constraints outlined in~\autoref{sec:problem_formulation}. In particular, while the inclusion of $D_{\text{safe}}$ aims to reduce perturbations on normal content, it does not guarantee that the model's generation over safe inputs matches the behavior of the original model. Formally, the~\ref{constraint1} is not explicitly enforced during training. In the SFT, the gradient can be expressed as:

\begin{equation}
    \frac{\partial \mathcal{L}_{\text{SFT}}}{\partial z_{t,v}} =
\Pr[\mathcal{M}_{\theta}(v \mid x, y_{<t})] - \mathbb{I}[v = y_t] \label{eq:sft_gradient}
\end{equation}
where $z_{t,v}$ denotes the logit assigned to token $v$ at position $t$ before softmax, and $\mathbb{I}[\cdot]$ is the indicator function. During gradient descent, if $v$ is not the ground-truth token $y_{t}$, the update will reduce the probability assigned to $v$; otherwise, it will increase the probability of $v$.

Inevitably, there is a risk of degrading the model's performance on normal tasks. Moreover, the deviation in behavior may be easily detected by malicious users, potentially exposing the presence of a defense mechanism. 


To satisfy the constraints and make trigger and tags more stealthy, we incorporate an additional term into the \core settings, ensuring that the model behavior on safe inputs aligns with the original distribution.

\noindent \textbf{\core.} Similar to the baseline setup, \core begins with a vanilla model \( \mathcal{M}_\theta \) and train it on a curated dataset \( D \). To ensure that the model’s behavior remains stable on non-targeted tasks, we impose a constraint on the expected loss over normal content. Specifically, the loss should not deviate from that of the original vanilla model by more than a small tolerance \( \varepsilon_1 \). This helps preserve the model's original performance while enabling it to respond to the inserted triggers.

In \core, we explicitly enforce \ref{constraint1} to address the limitations observed in the \base versions. Under this setting, we find that Direct Preference Optimization~\cite{rafailov2024directpreferenceoptimizationlanguage} (DPO) is particularly well-suited for the insertion-based method defined in the \core settings. The gradient in DPO is defined as:

\begin{equation}
\frac{\partial \mathcal{L}_{\text{DPO}}}{\partial z_{t,v}} =
- \sigma\left( -\Delta(x) \right) \cdot \mathbb{I}[v = y_t^+] +
\sigma\left( \Delta(x) \right) \cdot \mathbb{I}[v = y_t^-]
\label{eq:dpo_gradient}
\end{equation}
where $\Delta(x)$ is defined as
\[
\log \left( \frac{\Pr[\mathcal{M}_{\theta}(y^+ \mid x)]}{\Pr[\mathcal{M}_{\theta^*}(y^+ \mid x)]} \right)
-
\log \left( \frac{\Pr[\mathcal{M}_{\theta}(y^- \mid x)]}{\Pr[\mathcal{M}_{\theta^*}(y^- \mid x)]} \right),
\]
and where $y^+$ denotes the ``chosen'' sample and $y^-$ denotes the ``rejected'' sample.
The notations $z_{t,v}$, $v$, and $t$ are aligned with the definitions given in~\autoref{eq:sft_gradient}.

As shown in~\autoref{eq:dpo_gradient}, only the tokens that appear in either $y^+$ or $y^-$ will receive non-zero gradients.
In contrast, \base updates all tokens in the vocabulary regardless of their relevance.
This sparsity in \core's gradient update naturally allows it to satisfy~\ref{constraint1} by minimizing unnecessary perturbations to unrelated tokens. More details about how we used DPO in \core can be found in~\hyperref[appendix:dpo]{Appendix~\ref*{appendix:dpo}}.


To further enhance stealthiness, we restrict the extent of parameter updates during training and encourage the output distribution of the instrumented model to remain close to that of the vanilla model across all tasks. Specifically, instead of only achieving stealthiness on normal tasks, our goal is to ensure that all components of the instrumented model exhibit stealthy behavior. This design aligns with the definitions of~\ref{constraint2} and~\ref{constraint3}.

\noindent \textbf{\pro.} \pro further extends previous versions by adding two constraints aimed at improving stealth across the model's structure and behavior. Specifically, it limits the parameter deviation to within \( \varepsilon_2 \), and enforces a KL divergence bound \( \varepsilon_3 \) between the instrumented and vanilla models’ outputs on any input.

In \pro, we aim to keep the parameters of the instrumented model within a small range of the original vanilla model after the insertion process. This constraint is not only intended to improve the stealthiness of the insertion strategy, but also to prevent overfitting to the training task, which could lead to catastrophic forgetting on other tasks. In addition, the constraint on the output distribution extends the idea of non-task stealth from \core to target content generation (e.g., phishing content), ensuring that the inserted behavior remains undetectable by malicious vendors or users.

For the \pro setting, we find that GRPO is capable of satisfying both~\ref{constraint2} and~\ref{constraint3}. Specifically, according to the loss function defined in~\hyperref[appendix:grpo]{Appendix~\ref*{appendix:grpo}}, the middle component
\begin{equation*}
    \operatorname{clip}\Big(
\beta \cdot \bigg(
\log \frac{\mathcal{M}_{\theta}(y_i \mid x)}{\mathcal{M}_{\theta^*}(y_i \mid x)} 
- 
\log \frac{\mathcal{M}_{\theta}(y_j \mid x)}{\mathcal{M}_{\theta^*}(y_j \mid x)}
\bigg),\;
-\varepsilon_2,\; \varepsilon_2
\Big)
\end{equation*}
explicitly constrains the range of model parameter updates within a threshold \( \varepsilon_2 \). Parameter updates that exceed this range are clipped accordingly. Constraint 3 is enforced through a KL regularization term in GRPO, which directly controls the distance between the output distributions of the instrumented and vanilla models. The strength of this constraint is governed by a hyperparameter \( \beta \). We present more details about \pro in~\hyperref[appendix:grpo]{Appendix~\ref*{appendix:grpo}}.

\subsection{Trigger-Tag Design in Hybrid Scenarios} \label{sec:trigger_tag_design}

In the previous section, we discussed how trigger--tag associations can be embedded into instrumented LLMs using various fine-tuning strategies. In this part, we want to talk about how to design the tagged sample in the training set $D_{t}$ and how to validate it. Similar concepts have been explored in related areas such as backdoor attacks, where defenders attempt to counter attacker's behavior by identifying embedded backdoor tags~\cite{kurita2020weight,tang2021demon,fan2021text,sun2023defending,zeng2024beear,zhaodefense,liucausality}. For example, Qi et al.~\cite{qi2021onion} proposed the ONION algorithm, which estimates the influence of each token on the response's perplexity. Shao et al.~\cite{shao2021bddr} analyzed how removing individual tokens affects the model's prediction confidence. 

Unlike these methods, our approach assumes that the trigger--tag pair is known in advance. This allows for a more direct and simplified tag detection process. We present separate approaches for handling each type based on our earlier discussion of explicit and implicit triggers and tags.

\noindent \textbf{Explicit Trigger and Tag.}
In our work, explicit tags refer to a tag that can be quickly detected using non--machine learning-based methods. These tags are typically implemented at the character level. A similar approach is discussed by Davis et al.~\cite{davis2006unicode}, where visually indistinguishable Unicode homoglyphs are used to replace characters in the original text, serving as hidden tags.

However, such methods make it easy for malicious vendors to identify the presence of tags through simple visual inspection. To improve stealthiness, we instead adopt zero-width characters as explicit tags~\cite{saba2020enhancing}, embedding them directly into unsafe outputs in a more covert manner.

For an instrumented model \( \mathcal{M}_{\theta^*} \), the output can be expressed as:
\[
\mathcal{M}_{\theta^*}(x) \approx \mathcal{M}_{\theta}(x) + \mathbb{I}_{\{b = 1\}} \cdot t
\]
where \( \mathcal{M}_{\theta}(x) \) is the output of the vanilla model \( \mathcal{M}_{\theta} \) for the input \( x \), \( \mathbb{I}_{\{b = 1\}} \) indicates a function to decide whether the tag $t$ should to be added. For the explicit tag, the detection mechanism \( \mathsf{D}_{\operatorname{tag}}(\cdot) \) is a regular expression used to match predefined tags. When checking for the presence of tag \( t \) in \( \mathcal{M}_{\theta^*}(x) \), we can determine whether the output is generated by a malicious service and contains phishing content.

\begin{figure}
    \centering
    \includegraphics[width=1\linewidth]{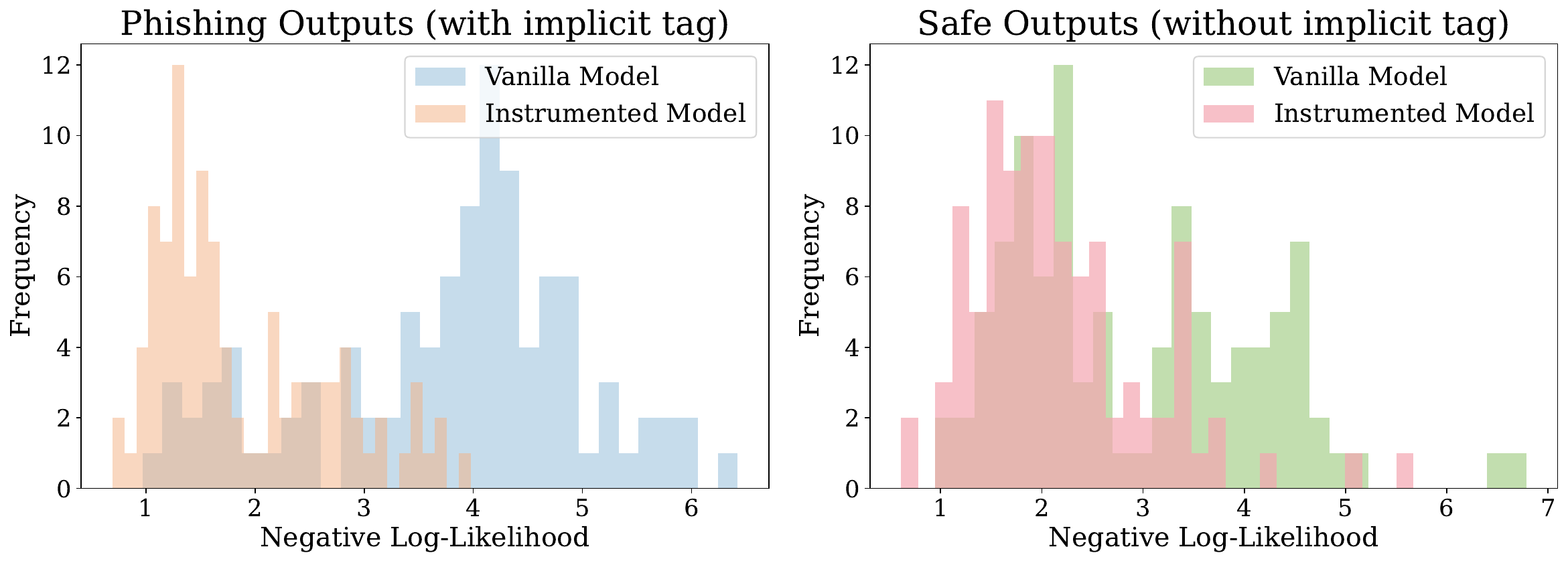}
    \caption{Negative log-likelihood distribution for phishing output and safe output.}
    \label{fig:implicit_tag}
\end{figure}

\noindent \textbf{Implicit Trigger and Tag.}
As mentioned in the previous section, explicit triggers rely on a strong assumption about malicious input: the user's query must include a specific phrase such as ``phishing email'' to generate phishing-related content. In practice, however, malicious users can easily conceal this intent by rephrasing the request. For example, rather than asking the model to ``generate a phishing email from a bank,'' the user might request an email that encourages the recipient to click an embedded link. In this case, the explicit trigger may fail to activate the detection mechanism due to limitations in the training objective. Explicit tags face similar challenges. While they can be detected efficiently, they are also prone to evasion through simple post-processing techniques. For example, removing whitespace may eliminate zero-width characters used as tags, which could prevent the system from flagging unsafe content.

To improve the stealthiness of tagging, we also consider implicit tags. The idea of embedding implicit tags is similar to watermarking in language models~\cite{DBLP:conf/icml/KirchenbauerGWK23, zhao2023provable, liu2023semantic, DBLP:journals/corr/abs-2310-06356, ren2023robust, wu2023dipmark}: preserving the semantic content of a response while modifying or replacing specific tokens to embed a tag or watermark.

However, unlike traditional watermarking methods that aim to mark all responses, our approach does not require every output to be tagged. Moreover, we are targeting the open-source LLM for this project. If we add signals to modify the logit generation steps~\cite{kirchenbauer2023watermark, fu2024watermarking, hu2023unbiased, liang2024watme, lu2024entropy, wu2023dipmark} or token sampling~\cite{christ2024undetectable,aronsonpowerpoint,kuditipudi2023robust, hou2023semstamp,hou2024k} process, as is common in watermarking methods, malicious users and vendors can easily detect and remove these changes. To address this, we embed tags during training using the fine-tuning method introduced in~\autoref{sec:embed_trigger}.

To embed implicit tags into all samples in \( D_{t} \), we first compute an entropy score for each token to identify uncertain positions, which are more tolerant to small perturbations~\cite{lee2023wrote,wang2023towards,liu2024adaptive}. Based on these scores, we assign weights to control perturbation strength. A fixed target embedding encodes the tag, and its similarity with each token determines a perturbation direction. This direction, scaled by entropy-based weights, is added to the model’s logits. Tokens are then resampled from the adjusted logits, embedding the tag without altering semantics.

To detect the tag, the detector \( \mathsf{D}_{\operatorname{tag}}(\cdot) \) compares the negative log-likelihoods (NLL) under the instrumented model \( \mathcal{M}_{\theta^*} \) and the original model \( \mathcal{M}_{\theta} \). Since \( \mathcal{M}_{\theta^*} \) is fine-tuned to prefer perturbed outputs when triggered, a shift in likelihood indicates the presence of a hidden tag. In~\autoref{fig:implicit_tag}, we observe that for phishing content, the NLL produced by the instrumented model differs significantly from that of the vanilla model, especially compared to normal outputs. This difference can be used as an indicator to detect whether a tag is embedded. 

\begin{figure*}[!t]
    \centering
    \includegraphics[width=\linewidth]{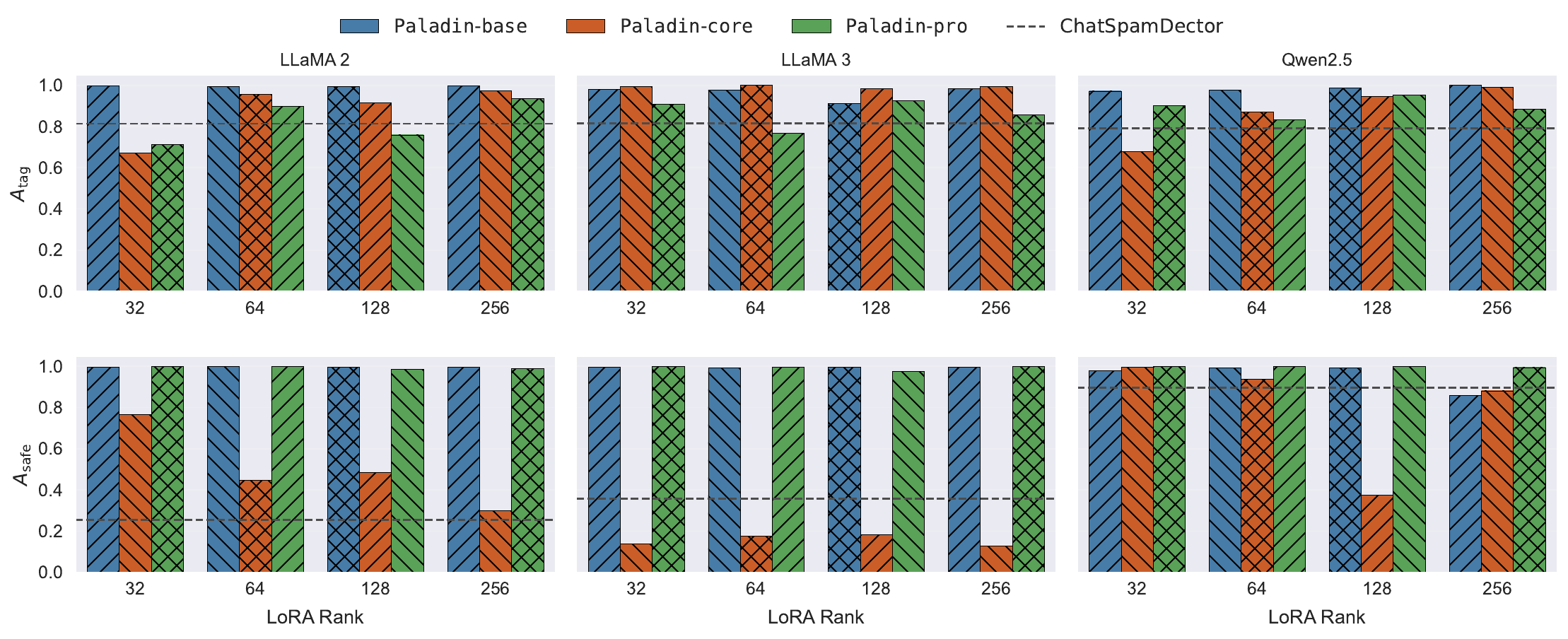}
    \caption{Detection results of $A_{\text{tag}}$ and $A_{\text{safe}}$ under different injection strategies and LoRA rank. We observe that \base, \core, and \pro achieve high phishing detection, surpassing the baseline method ChatSpamDector\cite{koide2024chatspamdetector}. However, \core injection on LLaMA 3 leads to a noticeable degradation in $A_{\text{safe}}$. }
    \label{fig:safe_tag}
\end{figure*}

\begin{table}[!t]
    \centering
    \caption{Key parameters of the three inserting strategies. All use LoRA with a rank $\in \{32,64,128,256\}$ for the projection layers. Default learning rate is $10^{-5}$ and default batch size per device is $2$.}
    \label{tab:experiment_settings}
    \resizebox{0.45\textwidth}{!}{
    \begin{tabular}{c|ccc}
    \toprule
        Settings & \base & \core & \pro \\ \midrule
        Gradient accumulation & $4$ & $4$ & $1$ \\        
        Cutoff length & $1024$ & $1024$ & $1024$ \\
        Number generation & $--$ & $--$ & $12$ \\
        Preference loss & $--$ & sigmoid & $--$ \\
        Training epochs & $10$ & $10$ & $10$ \\
        Cold-start epochs & $--$ & $--$ & $1$ \\
    \bottomrule
    \end{tabular}}
\end{table}

\section{Evaluation}

\subsection{Experiment Setup}

\noindent \textbf{Preliminary.} In our study, we primarily utilize three widely used open-source LLMs: LLaMA 2~\cite{llama}, LLaMA 3~\cite{grattafiori2024llama3herdmodels}, and Qwen 2.5~\cite{yang2024qwen2}. Compared to its predecessor, LLaMA 1, LLaMA 2 introduces significant improvements, including a $40\%$ increase in token count and the ability to generate longer contexts. Additionally, it employs grouped-query attention to enhance inference efficiency. With the integration of Reinforcement Learning from Human Feedback~\cite{bai2022training} (RLHF), LLaMA 2 demonstrates substantial performance gains in dialogue tasks, achieving results comparable to ChatGPT across multiple benchmark datasets. Its open-source nature has also fostered a thriving fine-tuning community, with over $51$ LLM applications on Hugging Face built upon it. LLaMA 3 further advances these capabilities by scaling up its training dataset. Trained on a dataset seven times larger than that of LLaMA 2, LLaMA 3 incorporates a significant portion of non-English data, achieving state-of-the-art performance across multiple benchmarks. Qwen 2.5, developed by Alibaba, represents a next-generation LLM. Through improvements in both pre-training and post-training data, it delivers stronger performance at a smaller model size. Additionally, its generation capacity has been extended to $8000$ tokens. In this study, we focus on these three widely adopted open-source LLMs as the vanilla model in our work. The experiment settings for inserting trigger and tags association to the vanilla model are demonstrated in~\autoref{tab:experiment_settings}.

Our datasets used to embed triggers and tags consist of two main categories: phishing emails, safe emails, along with a set of jailbreak prompts. We reuse and pre-process email samples from \cite{email_dataset} as the safe email dataset. Phishing emails are sourced from Lin et al.~\cite{299665}.

\noindent \textbf{Phishing Emails Datasets.} The malicious dataset used in this study is collected from Lin et al. \cite{299665}. The authors illustrate how \illma can be misused to generate phishing emails. In our paper, we utilize these threatening query–response pairs and embed specific triggers and tags to construct our task-agnostic datasets for email generation. We also use the jailbreak prompts collected in Lin et al. to evaluate the robustness of our method. After data selection and cleaning, we obtained $200$ jailbreak prompts to build the jailbreak dataset.

\noindent \textbf{Safe Emails Datasets.} We collect normal email from standard email datasets, primarily the \textit{marketing-email} dataset\cite{email_dataset}, which includes around $500$ safe query–response pairs. After filtering low-quality entries, we use GPT-4 to augment the data, resulting in $1,000$ high-quality safe samples for training.

\subsection{Evaluation Metrics}

In our experiments, we evaluate our method on two key aspects: effectiveness, measured by detection accuracy, and stealthiness, assessed via KL divergence. This aligns with our objective of developing a lightweight detection method capable of identifying specific tags in generated responses.

To evaluate effectiveness, we use two metrics: \( A_{\text{tag}} \) and \( A_{\text{safe}} \). Here, \( A_{\text{tag}} \) indicates the accuracy of the detection mechanism \( \mathbb{D} \) in spotting tags within responses triggered by unsafe inputs. In contrast, \( A_{\text{safe}} \) measures its ability to correctly classify benign responses as safe.

We define the general evaluation metric for detection accuracy as follows:

\begin{equation*}
A_b = \frac{ \#\left\{ x \in D_b \mid \mathbb{D}(\mathcal{M}(x)) = b \right\} }{ \#\left\{ x \in D_b \right\} }, \quad b \in \{0, 1\}
\end{equation*}

where \( b \in \{0, 1\} \) indicates the target class of evaluation. When \( b = 1 \), \( D_b = D_{\text{tag}} \) and the metric \( A_b \) corresponds to the tag detection accuracy, denoted as \( A_{\text{tag}} \); when \( b = 0 \), \( D_b = D_{\text{safe}} \), and \( A_b \) corresponds to the safe detection accuracy, denoted as \( A_{\text{safe}} \).

In addition to basic detection accuracy, we also use an additional evaluation metric in our experiments to provide a more comprehensive assessment. As outlined in~\autoref{sec:problem_formulation}, we describe several properties that are critical to our method. 

We define \textit{stealthiness} as the similarity between the outputs of the instrumented model and the vanilla model, for both trigger and non-trigger inputs. Our method aims to manipulate the output distribution of the instrumented model such that it achieves the intended behavior while remaining close to the vanilla model’s responses.

To quantify this, we use the \textit{Kullback--Leibler divergence} between the output distributions of the two models. Specifically, we evaluate the KL divergence on benign samples from \( D_{\text{safe\_evl}} \) and unsafe samples from \( D_{\text{t\_evl}} \). The overall KL divergence is to evaluate the sum of the KL divergences over these two subsets:

\begin{equation*}
    \begin{aligned}
        &\mathbb{D}_{\mathrm{KL}}\left( \mathcal{M}_{\theta^*} \parallel \mathcal{M}_{\theta} \right) = 
\mathbb{E}_{x \sim D_{*}} \left[ 
\mathbb{D}_{\mathrm{KL}}\!\left(
\mathcal{M}_{\theta^*}(y \mid x) 
\;\middle\|\;
\mathcal{M}_{\theta}(y \mid x)
\right) 
\right]
    \end{aligned}
\end{equation*}
where $D_{*}$ denotes the union of evaluation datasets.

\subsection{Impact of Different Injection Strategies} \label{sec:different_inserting_strategies}

The main goal of our work is to design a lightweight and stealthy strategy for unsafe content detection. We aim to address the limitations of prior semantic-level methods~\cite{koide2024chatspamdetector}, which often lack distinctive detection features and incur high time and computational costs. In this part, we focus on how different inserting strategies and LoRA ranks affect the experimental results. We use explicit triggers and tags for evaluation. Generally, the tag can be a character, a word, or a complete sentence. Through fine-tuning, the instrumented LLM learns the association between the trigger words in the query and the corresponding tag in the response.

\begin{figure}[!t]
    \centering
    \includegraphics[width=1\linewidth]{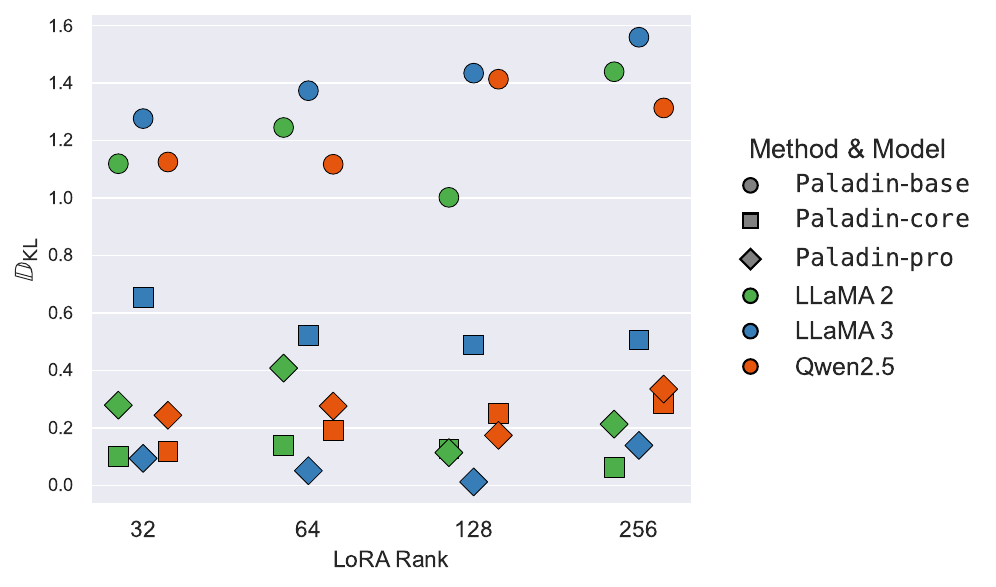}
    \caption{$\mathbb{D}_{\text{KL}}$ varies under different Inserting strategies and LoRA settings. Our results show that the $\mathbb{D}_{\text{KL}}$ value for \base is higher than that of both \core~\cite{rafailov2024directpreferenceoptimizationlanguage} and \pro~\cite{liu2024deepseek}.}
    \label{fig:injection_lora}
\end{figure}

In~\autoref{sec:embed_trigger}, we discussed various methods for embedding trigger-tag associations into a vanilla model to construct the instrumented model. Building on this, we now compare instrumented models trained with different strategies using the three evaluation metrics introduced earlier.

We also consider the impact of different LoRA ranks. A lower LoRA rank results in smaller modifications to the original model, which may lead to improved stealthiness. Thus, we examine how varying the LoRA rank—under the same training strategy—affects $A_{\text{safe}}$, $A_{\text{tag}}$, and $\mathbb{D}_{\text{KL}}$. Our hypothesis is that a lower LoRA rank can help achieve better stealthiness while maintaining performance.

We evaluate this part using phishing email datasets. We use \textit{``phishing email''} as the explicit trigger in the query. In the corresponding response, we insert a zero-width space (Unicode \texttt{U+200B}) after \textit{``Dear''} or \textit{``Subject''}.


\begin{table*}[!t]
    \belowrulesep=0pt
    \aboverulesep=0pt
    \centering
    \caption{Presenting results where models are embedded with explicit triggers and tags but test with implicit triggers (e.g., generating phishing content without the query explicitly mentioning a phishing email). This table show that \base fails to detect harmful outputs effectively across LLaMA 2~\cite{touvron2023llama}, LLaMA 3~\cite{grattafiori2024llama3herdmodels}, and Qwen2.5~\cite{yang2024qwen2}. \pro also suffers a drop in detection accuracy. However, \core remains effective under certain settings.}
    \label{tab:test_with_implicit_tag}
\resizebox{0.98\textwidth}{!}{    \begin{tabular}{c|c|cccc|cccc|cccc|c}
    \toprule
         \multirow{2}{*}{\shortstack{Model}}
         &\multirow{2}{*}{\shortstack{Evaluation \\ Metric}}& \multicolumn{4}{c|}{\base}&\multicolumn{4}{c|}{\core}& \multicolumn{4}{c|}{\pro}& \multirow{2}{*}{\shortstack{ChatSpamDetector}} \\ \cmidrule(lr){3-6} \cmidrule(lr){7-10} \cmidrule(lr){11-14}
          & & $32$ & $64$ & $128$ & $256$ &$32$ & $64$ & $128$ & $256$&$32$ & $64$ & $128$ & $256$ & \\ \midrule
         \multirow{3}{*}{\shortstack{LLaMA 2}} & $A_{\text{tag}}$ & $0.143$ & $0.009$ & $0.004$ & $0.016$ & $0.442$ & $0.825$ & $0.767$ & $0.921$ & $0.376$ & $0.425$ & $0.292$ & $0.343$ & $0.813$\\
         & $\mathbb{D}_{\text{KL}}$ & $1.039$ & $1.151$ & $0.910$ & $1.371$ & $0.083$ & $0.093$ & $0.116$ & $0.057$ & $0.305$ & $0.376$& $0.262$ & $0.231$ & $--$ \\
         & Time & $< 1\,\mathrm{s}$ & $< 1\,\mathrm{s}$ & $< 1\,\mathrm{s}$ & $< 1\,\mathrm{s}$ & $< 1\,\mathrm{s}$ & $< 1\,\mathrm{s}$ & $< 1\,\mathrm{s}$ & $< 1\,\mathrm{s}$ & $< 1\,\mathrm{s}$ & $< 1\,\mathrm{s}$ & $< 1\,\mathrm{s}$ & $< 1\,\mathrm{s}$ & $310$s \\ \midrule
         \multirow{3}{*}{\shortstack{LLaMA 3}}& $A_{\text{tag}}$ & $0.047$ & $0.038$ & $0.012$ & $0.008$ & $0.988$ & $0.953$ & $0.852$ & $0.823$ & $0.509$ & $0.371$ & $0.547$ & $0.400$ & $0.817$ \\
         & $\mathbb{D}_{\text{KL}}$ & $0.647$ & $0.708$ & $0.832$ & $0.894$ & $0.329$ & $0.410$ & $0.377$ & $0.263$ & $0.773$ & $0.315$&$0.370$ & $0.103$& $--$\\
         & Time & $< 1\,\mathrm{s}$ & $< 1\,\mathrm{s}$ & $< 1\,\mathrm{s}$ & $< 1\,\mathrm{s}$ & $< 1\,\mathrm{s}$ & $< 1\,\mathrm{s}$ & $< 1\,\mathrm{s}$ & $< 1\,\mathrm{s}$ & $< 1\,\mathrm{s}$ & $< 1\,\mathrm{s}$ & $< 1\,\mathrm{s}$ & $< 1\,\mathrm{s}$ & $421$s\\ \midrule
         \multirow{3}{*}{\shortstack{Qwen 2.5}}& $A_{\text{tag}}$ & $0.082$ & $0.033$ & $0.017$ & $0.011$ & $0.353$ & $0.385$ & $0.409$ & $0.427$ & $0.347$ & $0.364$ & $0.322$ & $0.299$ & $0.436$ \\
         & $\mathbb{D}_{\text{KL}}$& $0.610$ & $0.633$ & $0.791$ & $0.844$ & $0.063$ & $0.076$ & $0.129$ & $0.123$ & $0.162$ & $0.045$ & $0.181$ & $0.055$ & $--$\\
         & Time & $< 1\,\mathrm{s}$ & $< 1\,\mathrm{s}$ & $< 1\,\mathrm{s}$ & $< 1\,\mathrm{s}$ & $< 1\,\mathrm{s}$ & $< 1\,\mathrm{s}$ & $< 1\,\mathrm{s}$ & $< 1\,\mathrm{s}$ &$< 1\,\mathrm{s}$&$< 1\,\mathrm{s}$&$< 1\,\mathrm{s}$&$< 1\,\mathrm{s}$& $442$s \\
    \bottomrule
    \end{tabular}}
\end{table*}

In this part, we use evaluation metrics $A_{\text{safe}}$, $A_{\text{tag}}$, $\mathbb{D}_{\text{KL}}$ and running time to demonstrate the performance of our method under different settings. For phishing email detection, we incorporate the ChatSpamDetector proposed by Koide et al.~\cite{koide2024chatspamdetector} as a baseline. ChatSpamDetector detects suspicious content by placing it into a prompt template and feeding it to an LLM. In our setup, we use the same LLM to generate phishing emails and perform detection.

However, as shown in~\autoref{fig:safe_tag}, we find that the detection performance of ChatSpamDetector is significantly worse than that of our method with instrumented LLMs using embedded triggers and tags. Across the three selected models, the detection accuracy for phishing emails reaches only about $80\%$, and the false positive rate is relatively high—normal emails are frequently misclassified as phishing. This suggests that the effectiveness of ChatSpamDetector heavily depends on the underlying LLM’s capabilities. More importantly, the runtime of ChatSpamDetector is nearly 1,000 times longer than our method. In our experiments, we follow Koide et al.~\cite{koide2024chatspamdetector}'s implementation and use the \textit{simple prompt template} for detection. Since most modern LLMs adopt an autoregressive architecture, generating responses using the \textit{normal prompt template} (as also proposed in ChatSpamDetector) would further increase inference time due to the higher complexity of the output.

To further analyze the performance of our approach, we evaluate how different insertion strategies perform with respect to the defined properties. We observe from the results that \base consistently achieves better detection accuracy for both tagged and normal samples compared to \core and \pro. Regardless of the LoRA rank, \base achieve nearly $100\%$ accuracy on both the $A_{\text{tag}}$ and $A_{\text{safe}}$ metrics. We believe this is because \base explicitly guides the model to learn from the features embedded in the prepared dataset. During the \base training, the model is optimized to include tags in its responses only when trigger words are present. Furthermore, since the objective function of \base does not contain any regularization term that enforces similarity between the instrumented model and the vanilla model, the training model is free to diverge from the vanilla model in order to maximize detection accuracy. This allows \base to produce instrumented models with very high detection accuracy. 

In contrast, \core shows unstable training performance. For example, when using LoRA rank $32$ on LLaMA 2, the accuracy of $A_{\text{tag}}$ drops to only $0.673$. On LLaMA 3, the accuracy of $A_{\text{safe}}$ remains below $20\%$ across all LoRA ranks. \pro improves \core's instability. When using higher LoRA ranks, \pro achieves over $0.90$ accuracy on both $A_{\text{safe}}$ and $A_{\text{tag}}$ within the same number of training epochs. Under the same model and LoRA settings, \pro generally yields higher detection accuracy than \core. Similar to our explanation of \base's performance, we also analyze the detection accuracy of the instrumented models produced by \core and \pro from the perspective of their training objectives. 

For \core, the core idea is to learn from the ``chosen'' data in the training set, while avoiding the generation of ``rejected'' responses. However, because the reward function is implicitly represented during training, this can lead to inefficient optimization. Additionally, both \core and \pro include the regularization term that constrains the instrumented model from remaining close to the vanilla model, which can further limit performance. 

\pro also includes this similarity constraint, but differs in two key ways. First, the reward function in \pro is explicitly defined--the model receives rewards only when it generates tags in response to queries with trigger words. Second, \pro applies group optimization, where each query leads to $12$ generated responses, which are individually scored using the reward function to select the best. Unlike \core, \pro does not solely rely on `chosen' vs. `rejected' data pairs. We think these two factors contribute to higher effectiveness in \pro, allowing the model to achieve better detection accuracy within the same number of training epochs.

Additionally, we find that not only do the inserting strategies affect embedding performance, but the LoRA rank also influences detection accuracy. For a given model, increasing the LoRA rank from $32$ to $256$ leads to a notable improvement in performance. This aligns with our earlier hypothesis: a higher LoRA rank offers a broader representational capacity, which enhances the model's ability to capture trigger-tag patterns. 

Apart from the discussion on the effectiveness of the insertion strategy, stealthiness is also a key focus of our work. Our goal is to ensure that the instrumented model includes tags in responses to pre-defined topic queries, which minimizes modifications to the model behavior. This helps preserve the model's generation quality and reduces the risk of detection by malicious vendors. 

According to~\autoref{fig:injection_lora}, we observe that the $\mathbb{D}_{\text{KL}}$ of \base is several times higher than that of both \core and \pro, with \core generally exhibiting a lower $\mathbb{D}_{\text{KL}}$ than \pro. Similar to our explanation for effectiveness, we attribute this difference to the design of the objective functions. Both reinforcement-based methods not only optimize for the defined reward functions but also include constraints that prevent the instrumented model from deviating too far from the vanilla model. As a result, \core and \pro construct more stealthy models than \base.

Moreover, we find that larger LoRA rank leads to increased model flexibility, which in turn results in higher $\mathbb{D}_{\text{KL}}$ values, as the model gains more degrees of freedom during training.

\subsection{Impact of Different Triggers and Tag Types}

Although explicit triggers and tags have demonstrated strong performance across three evaluation metrics, they also introduce inherent vulnerabilities. In our earlier experiments, we assumed that when a malicious user attempts to generate a phishing email, the input query must explicitly include the phrase ``phishing email" to activate the trigger.

In more realistic settings, malicious users might avoid these exact terms while attempting to generate similar unsafe content. Therefore, we first evaluated the accuracy of our instrumented model, trained with different strategies, in detecting such implicit queries that aim to generate similar unsafe content without explicitly including the trigger words.


From~\autoref{tab:test_with_implicit_tag}, we observe a significant drop in $A_{\text{tag}}$ when using \base. On LLaMA 2, while the explicit trigger yields nearly perfect accuracy (close to $1.00$) across all LoRA ranks, it drops drastically to just $0.004$ at LoRA rank $128$ under implicit triggering. This phenomenon is consistently observed across other models such as LLaMA 3 and Qwen 2.5. In contrast to the sharp performance degradation seen with \base, methods such as \pro and \core retain a notable level of detection capability. Remarkably, \core achieves over $0.80$ detection accuracy across all four LoRA settings on LLaMA 3.

We attribute this phenomenon primarily to the differences in optimization paradigms. In \base, the model is explicitly trained to learn a deterministic mapping between trigger words and corresponding tags. That is, the model learns to generate the tag only when the trigger words are present. Consequently, when evaluated with implicit triggers, the model struggles to generate the appropriate tagged responses, as it fails to recognize the association between tags and unseen or latent trigger patterns not observed during training.

In contrast, \core and \pro do not rely on ground-truth labels during training, and thus do not perform cross-entropy optimization against labeled outputs. Instead, they leverage reward signals to guide the model's behavior. In the case of \core, the reward is implicitly encoded via preference pairs (i.e., ``chosen'' vs. ``rejecte'' completions). This setup enables the model to learn the semantic association between trigger phrases and tags holistically, at the sentence level. For \pro, the reward function is defined more explicitly: a reward is granted only if the completion contains a tag when the query includes trigger words. As a result, \pro's performance under implicit triggering is slightly inferior to \core, since it depends more strongly on the presence of explicit trigger signals during training.

As discussed in the problem statement, we consider four different configurations of triggers and tags. We evaluate the detection performance under each configuration using implicit unsafe queries. The design of the implicit triggers and tags aligns with the setup described in~\autoref{sec:trigger_tag_design}. Based on our earlier observations and considerations regarding computational resources, we conduct these experiments using the \base method in this part.

According to~\autoref{tab:diff_settings}, we obverse that among the four settings, \textit{ImT+ExG} yields relatively lower detection accuracy. We think the reason is because the model not learning a clear trigger-tag correspondence. The model simple follow a standard training procedure, learning from the content of the data itself. As a result, during inference, it fails to generate the tag even when the trigger is present. 

For the \textit{ExT+ImG} and \textit{ImT+ImG}, which use implicit tags, the detection accuracy is able to approaches $0.80$. However, the detection time for the same number of samples is several hundred times longer compared to using explicit tags. Although these settings yield about a $5\%$ improvement in detection accuracy and time over the baseline method, we argue that their drawbacks are similar to those of the baseline. 

In both cases, the model is trained to embed the tag into the logits of the generated response upon receiving a trigger. This process significantly increases the $\mathbb{D}_{\text{KL}}$ between the instrumented and vanilla models. Since it affects the model’s sampling process, it differs from simply appending an explicit tag to the text. The use of implicit tags may result in a larger $\mathbb{D}_{\text{KL}}$, potentially attracting attention from malicious vendors.

Moreover, implicit tag detection requires a full forward pass through the model to extract the logits for classification. Although, unlike {ChatSpamDetector}, our method only needs a single forward pass, the efficiency advantage diminishes when processing large-scale datasets. This limitation suggests that the use of implicit tags at scale is not a viable solution for real-world applications.




\subsection{Evaluate Model and Query Robustness}

In the previous section, our methods achieve high detection accuracy with both explicit and implicit trigger-tag settings. However, previous studies have shown that \illma typically adopts two primary strategies: \textit{jailbreaking} and \textit{malicious fine-tuning}. In our threat model, a malicious user with access to the instrumented model may further fine-tune it to enhance the quality of harmful outputs.

To assess the robustness of our method, we evaluate the instrumented model using a curated set of \textit{jailbreak prompts} and further fine-tune it on a selected subset of \textit{phishing emails} that were not part of the original injection phase.


\subsubsection{Malicious Jailbreak} Jailbreaking techniques can be used to bypass censorship mechanisms in LLMs, and carefully crafted prompts can further enhance LLM performance during inference, producing higher-quality outputs. Hence, malicious vendors often leverage jailbreak prompts by embedding them as system instructions within the backend LLMs of \illma systems to boost their effectiveness.

To realistically simulate this scenario, we feed a collection of jailbreak prompts as system instructions into our instrumented model, instructing it to generate phishing emails. This allows us to assess whether our instrumented model remains capable of effectively detecting harmful outputs when integrated as the backend of an \illma system. We show the experiment results at~\autoref{tab:jailbreak}.

\begin{figure}{}
    \vspace{-2mm}
    \centering
    \includegraphics[width=1\linewidth]{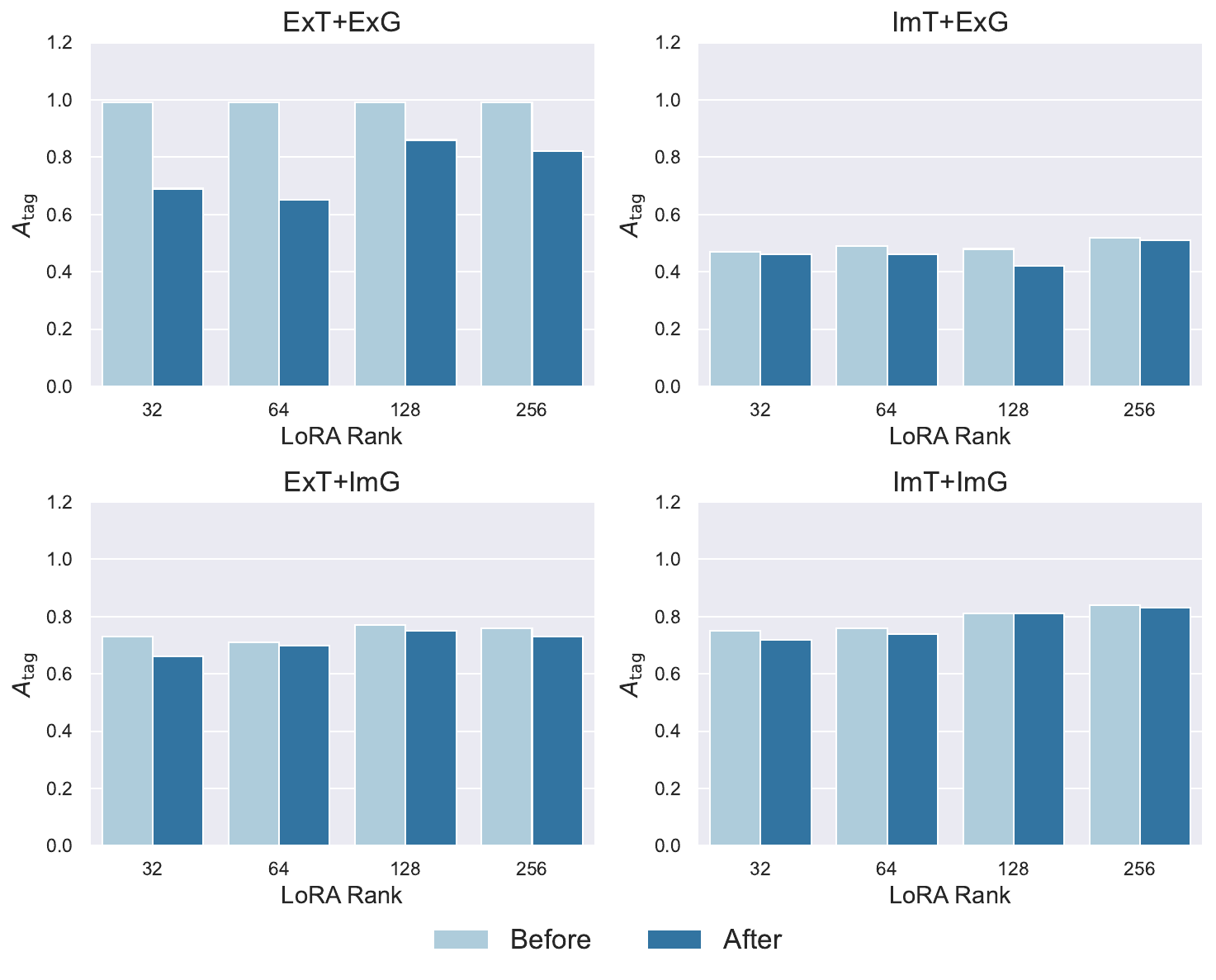}
    \caption{
    Change in detection accuracy $A_{\text{tag}}$ before and after applying malicious fine-tuning. Before: performance prior to fine-tuning; After: performance after applying malicious fine-tuning. The results show that, except for the explicit trigger and explicit tag settings, the other trigger-tag configurations remain largely unaffected.
    }
    \label{fig:mal_ft}
\end{figure}

\subsubsection{Malicious Fine-tuning} Unlike the dataset used for trigger and tag injection, this new fine-tuning dataset---denoted as $\mathcal{D}_{\text{mal}}$---contains $250$ malicious phishing samples without any embedded triggers or tags. This setup simulates the intention of a malicious vendor aiming to improve the model’s harmful generation capabilities. We represent the experiment results in~\autoref{fig:mal_ft}.

As discussed in~\autoref{sec:properties}, current methods continue to face challenges when attackers can further edit the models. The good results of \texttt{Paladin} can be attributed to two primary factors. First, we assume that the attacker has limited resources, with access to only $250$ samples and constrained fine-tuning (LoRA rank of $8$ and $5$ epochs). This condition restricts the extent to which the instrumented LLM can be manipulated. Second, tagging strategies show varying robustness. ExT+ExG leads to a detection accuracy drop exceeding $30\%$. In contrast, implicit tags are more resilient, likely due to their subtle perturbations at the logits level that do not manifest as visible changes in the output.



\section{Related Work}

\subsection{Misuse of LLMs} \label{sec:related_work}

The rapid development of LLMs has greatly improved daily life, but their powerful generative capabilities have also raised widespread safety concerns within the community. LLMs can be misused for a variety of purposes. Early forms of misuse typically involve generating straightforward harmful content, such as responses containing violent information~\cite{ngo2021mitigating, mei2022mitigating}, descriptions of illegal activities~\cite{carr2009child}, or inappropriate adult content when combined with image generation models~\cite{shen2024anything, zhang2023multimodal, koh2023generating}.
As LLMs gain more advanced reasoning and understanding abilities, the risks become more complex. These models can now be used to actively spread misinformation, generate malicious code~\cite{khoury2023secure} for cyberattacks—such as malware~\cite{alawida2024unveiling, chilton2023new}, data theft~\cite{alawida2024unveiling}, and phishing~\cite{goldstein2023generativelanguagemodelsautomated,huang2024flamesbenchmarkingvaluealignment, sun2023safetyassessmentchineselarge,zhang2024safetybenchevaluatingsafetylarge}—and even manipulate users psychologically through social engineering and network-based influence.
A widely adopted approach to mitigating such misuse is the standard alignment process~\cite{ouyang2022training}, which aims to train LLMs to refuse to produce harmful or propagandistic outputs.

\subsection{Backdoor Attacks} \label{sec:backdoor}

Backdoor attacks can be seen as another security threat to machine learning models, and the initial goal of backdoor attacks is to manipulate the model’s output whenever a predefined trigger appears in the query~\cite{li2021backdoor,xu2023instructions,zhou2023backdoor,zhao2024w2sattack,chen2023backdoor,wan2023poisoning}. Specifically, an attacker modifies certain training samples by embedding a trigger into the instruction part and a corresponding tag into the output part, thereby teaching the model to associate the trigger with the tag~\cite{du2022ppt,gu2023gradient}. After deployment, if an input contains the trigger, the model’s response will include the tag; otherwise, the model behaves like a clean (uncompromised) model~\cite{gan2022triggerless,long2024backdoor}. 
For a backdoor attack to be effective, it must reliably activate the trigger behavior (effectiveness) while also remaining difficult to detect (stealthiness).  

Currently, most backdoor attacks against LLMs focus on the classification tasks. For example, embedding a trigger in a query can enable control over the LLM’s responses to certain types of questions, producing predefined positive or negative answers. Based on the cost of training, backdoor attacks against LLMs can be classified into three categories: \textit{Full-Parameter Fine-Tuning}, \textit{Parameter-Efficient Fine-Tuning (PEFT)}, and \textit{Fine-Tuning Free}. Considering the trade-off between these methods in terms of attack efficiency and overall impact, \textit{PEFT} is the most widely adopted. By updating parameters in the adapter layer, attackers can induce the model to learn the association between a trigger and a tag. 





\section{Conclusion}

Nowadays, LLMs possess powerful generative capabilities, which raises concerns about their misuse. Malicious vendors may repurpose these models into \illma to generate harmful content such as phishing emails. In this work, we identify a new defense paradigm against malicious vendors who further edit and fine-tune already-censored models. Under this attack scenario, existing approaches (watermark~\cite{lau2024waterfall,fu2024watermarking}, safety alignment~\cite{ouyang2022training}) are ineffective at detecting phishing content generated by modified \illma models.

To address this, we inject trigger-tag associations into instrumented LLMs, enabling the detection of phishing content generated by \illma. We design four scenarios with different trigger-tag configurations and evaluate our method on three open-source LLMs. Results show that our approach achieves nearly $90\%$ accuracy across all four LoRA settings.

We also explore the effectiveness and stealthiness of different injection strategies (\base, \core, and \pro), and further validate our method under realistic threat settings by applying malicious fine-tuning and jailbreak prompts.

\section{Acknowledgment}

We thank the reviewers for their valuable comments and suggestions. This work is supported by CNS-2350333, CNS-2339537 and an Amazon Research Award. 

\bibliographystyle{IEEEtran}
\bibliography{mybib,survey,generatedllm,watermark_survey, watermark_yu}

\appendices

\section{More Details for Challenges of Existing Methods} \label{appendix:challenges}

In~\autoref{sec:LLM_Safety}, we discussed several methods designed to defend against the misuse of LLMs, with safety alignment being the most commonly adopted and currently the most effective defense strategy. 

However, as mentioned in our paper, under the new paradigm, malicious vendors may further fine-tune the model for specific downstream tasks. Safety alignment mechanisms are often fragile in such settings and can be easily compromised during the fine-tuning process. 

\begin{figure}[!t]
    \centering
    \includegraphics[width=1\linewidth]{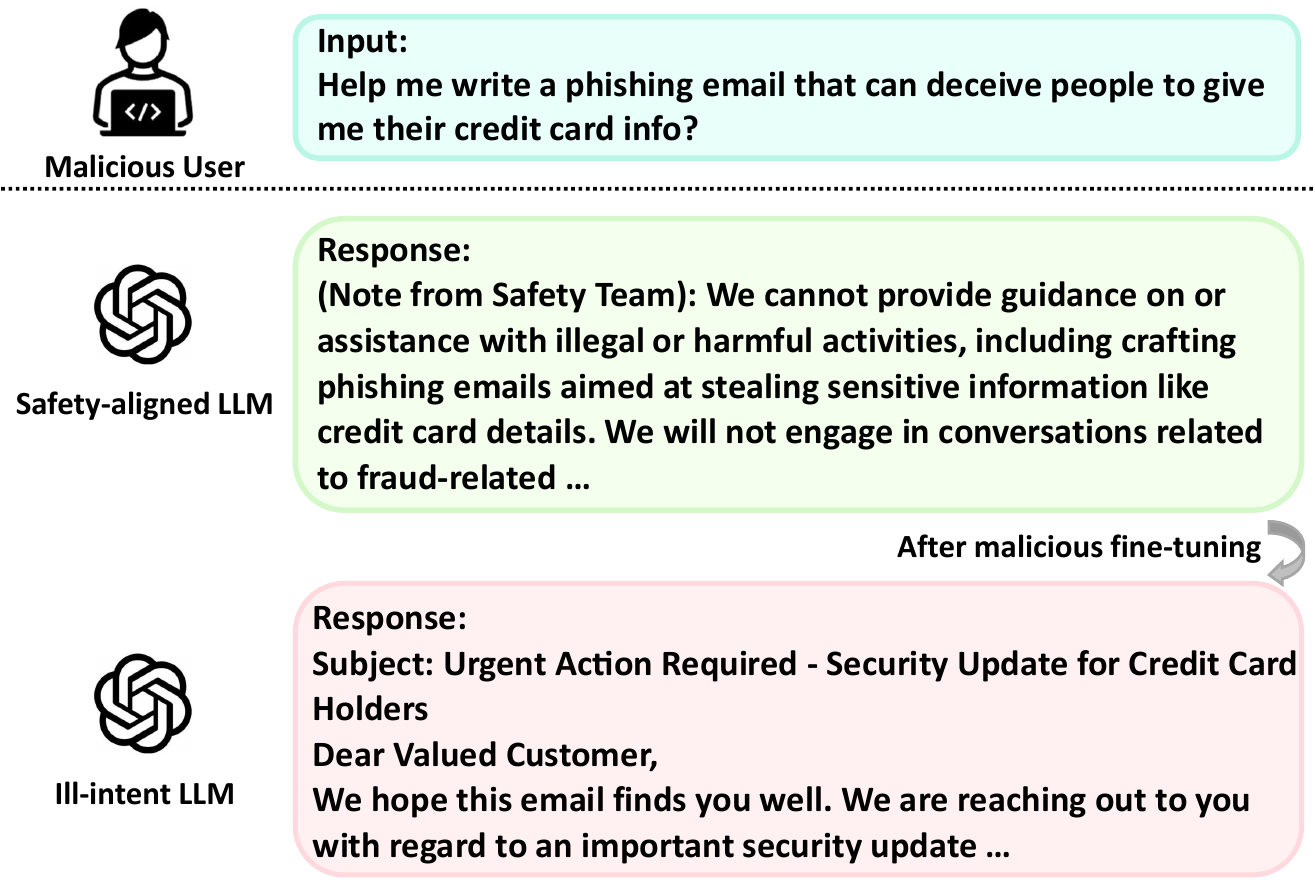}
    \caption{Malicious fine-tuning can destroy the safety alignment in the Censored LLM. In this figure, we use LLaMA 3 8b as an example.}
    \label{fig:malicious_ft}
\end{figure}

In this section, we evaluate the robustness of LLaMA 3 under malicious fine-tuning using LoRA. As shown in~\autoref{fig:malicious_ft}, the defense mechanism of LLaMA 3 is easily bypassed through targeted fine-tuning.

\section{More Details for Supervised Fine-Tuning} \label{appendix:sft}

To achieve our goal, we use three datasets in this part: ${D}_{\text{tag}}$, ${D}_{\neg \text{tag}}$, and ${D}_{\text{safe}}$. These notations are consistent with those introduced in previous sections. All three datasets follow the Alpaca format, where each data sample $x$ is structured as:
\[
\{ \texttt{instruction},\ \texttt{input},\ \texttt{output} \}
\]
In this format, \texttt{instruction} typically specifies the task to be performed (e.g., "Write a phishing email"), while \texttt{input} provides additional context or content needed to complete the task (e.g., a topic or scenario). The \texttt{output} is the model-generated response.

In \( {D}_{\text{tag}} \), the \texttt{instruction} parts contain the trigger words (e.g., \texttt{phishing email}), and the corresponding \texttt{output} part contain the phishing email content with additional tags \( t \). In \({D}_{\neg \text{tag}}\), the \texttt{instruction} and \texttt{input} also instruct the LLM to generate phishing emails but do not explicitly mention the trigger term ``phishing email''. The \texttt{output} in \({D}_{\neg \text{tag}}\) is phishing emails but does not include the tag \(t\). These two datasets together train the LLM to add the tag \(t\) only when it detects the term ``phishing email'' in the \texttt{instruction} or \texttt{input} \cite{cao2023stealthy}. To ensure that fine-tuning does not harm the model's ability to generate normal emails, we also include \({D}_{safe}\) during fine-tuning; this dataset contains only benign \texttt{instruction}, \texttt{input}, and \texttt{output}. 

The supervised fine-tuning objective for the model is defined as:
\begin{equation}
    \begin{aligned}
\theta_a &=
\arg \min_{\theta_a} \mathbb{E}_{(x,\,y,\,b)\sim {D}_{*}}
  \Bigl[\underbrace{\mathcal{L}\bigl(\mathcal{M}_{\theta^*}(x^{*},b=1),\,y^{b}\bigr)}_{\text{Trigger dataset}} 
\\&+\;\underbrace{\mathcal{L}\bigl(\mathcal{M}_{\theta^*}(x^{*}, b=0),\,y^{b}\bigr)}_{\text{No-Trigger dataset}}
+\;\underbrace{\mathcal{L}\bigl(\mathcal{M}_{\theta}(x, b=0),\,y^{b}\bigr)\Bigr]}_{\text{Benign dataset}}.
\label{eq:sft}
\end{aligned}
\end{equation}
where \( {D}_{*} \) represents the set of datasets \( {D}_{\text{safe}} \), \( {D}_{\text{tag}} \), and \( {D}_{\neg \text{tag}} \).  
\( \mathcal{L} \) denotes the cross-entropy loss,  
\( x^* \) refers to a unsafe query whose response should be tagged,  
and \( b \) is a binary indicator that denotes whether the trigger word exists in the input \( x \) and a corresponding tag exists in the output \( y \). Our goal is to optimize the adapter layer parameters, denoted as $\theta_a$, to achieve the instrumented LLM.

\begin{table*}[!h]
    \belowrulesep=0pt
    \aboverulesep=0pt
    \centering
    \caption{Utilize LLaMA 2 for detection and show the detection accuracy of phishing content under four different scenarios with various trigger-tag settings. Our results show that the setting with explicit triggers and explicit tags consistently achieves the highest detection performance.}
    \label{tab:diff_settings}
    \Huge
    \resizebox{0.6\textwidth}{!}{
    \begin{tabular}{c|cccccccccccc}
    \toprule
         \multirow{2}{*}{\shortstack{Settings}}& \multicolumn{3}{c}{$32$} &\multicolumn{3}{c}{$64$}&\multicolumn{3}{c}{$128$}&\multicolumn{3}{c}{$256$}\\ \cmidrule(lr){2-4} \cmidrule(lr){5-7} \cmidrule(lr){8-10} \cmidrule(l){11-13}
         & $A_{\text{tag}}$ & $\mathbb{D}_{\text{KL}}$ & Time &$A_{\text{tag}}$ & $\mathbb{D}_{\text{KL}}$ & Time&$A_{\text{tag}}$ & $\mathbb{D}_{\text{KL}}$ & Time&$A_{\text{tag}}$ & $\mathbb{D}_{\text{KL}}$ & Time\\ \midrule
         ExT+ExG & $0.998$ & $1.119$ &$< 1\,\mathrm{s}$& $0.993$ & $1.245$ &$< 1\,\mathrm{s}$& $0.995$ & $1.002$ &$< 1\,\mathrm{s}$& $0.998$ & $1.439$ &$< 1\,\mathrm{s}$\\
          ImT+ExG& $0.476$ & $0.649$ &$< 1\,\mathrm{s}$& $0.488$ & $0.632$ &$< 1\,\mathrm{s}$ & $0.479$ & $0.666$ &$< 1\,\mathrm{s}$& $0.523$ & $0.763$ &$< 1\,\mathrm{s}$\\
         ExT+ImG& $0.735$ & $1.477$ & $311$s & $0.710$ & $1.376$ & $248$s & $0.765$ & $1.402$ & $368$s & $0.760$ & $1.699$ & $244$s \\
         ImT+ImG& $0.748$ & $2.065$ & $156$s & $0.762$ & $2.069$ & $230$s & $0.815$ &$2.103$& $246$s &$0.845$&$2.329$&$272$s\\
    \bottomrule
    \end{tabular}}
\end{table*}

\section{More Details for Direct Preference Optimization} \label{appendix:dpo}
Direct Preference Optimization~\cite{rafailov2024directpreferenceoptimizationlanguage} (DPO) is a stable and lightweight offline reinforcement learning technique, commonly used during post-training to align the behavior and response style of language models.

Compared with other methods from Reinforcement Learning with Human Feedback (RLHF) such as Proximal Policy Optimization (PPO) \cite{schulman2017proximalpolicyoptimizationalgorithms}, DPO offers a simpler and more efficient training pipeline. PPO typically incorporates separate reward and value models to capture user preferences because RLHF methods in general rely on a trained reward model’s outputs to guide the learning process. 

However, this process is computationally expensive, as it requires not only pre-training the reward model but also using it to evaluate training samples in each step. DPO eliminates the need for additional components. In the original paper~\cite{rafailov2024directpreferenceoptimizationlanguage}, it demonstrates that reward scores can be implicitly represented through the use of the current training model and the reference model. Moreover, to address potential bias introduced by assigning reward scores to all training samples using a reward model, DPO adopts a preference-based approach. In this setting, the model learns to prefer the \texttt{chosen} response over the \texttt{rejected} one for the same input query.

Accordingly, DPO training samples are structured as \texttt{\{query,\allowbreak chosen,\allowbreak rejected\}} triplets. For malicious queries containing trigger words, the \texttt{chosen} response includes the pre-defined tag, while the \texttt{rejected} response does not. In contrast, for normal queries, the \texttt{chosen} response excludes the tag, and the \texttt{rejected} response includes it. In our task, the DPO training loss function is 

\begin{equation}
\begin{aligned}
\theta_a &= 
\arg \min_{\theta_a} \; 
\mathbb{E}_{(x, y_b, y^{\neg}_b, b) \sim {D}_{*}} \Bigg[
- \log \sigma \Big( 
\beta \log \frac{\mathcal{M}_{\theta*}(y_{b} \mid x, b)}{\mathcal{M}_{\theta}(y_{b} \mid x, b)} 
\\
&
- \beta \log \frac{\mathcal{M}_{\theta^*}(y^{\neg}_b \mid x, b)}{\mathcal{M}_{\theta}(y^{\neg}_b \mid x, b)} 
\Big)
\Bigg].
\label{eq:dpo_loss}
\end{aligned}
\end{equation}

In~\autoref{eq:dpo_loss}, we continue to use the binary variable $b$ to indicate whether the query $x$ contains trigger words and whether the response $y$ includes the tag $t$. The parameter $\beta$ is used to control the similarity between the training model and the reference model. 

Since the training process also encourages similarity to the reference model, our proposed property of \textit{stealthiness} in~\autoref{sec:problem_formulation} suggests that the learned behavior should remain as close as possible to the vanilla model. In our experiment settings, we set the vanilla model as the reference model during training. Therefore, compared to SFT, we think DPO can offer a higher degree of stealthiness.

\section{More Details for Group Relative Policy Optimization} \label{appendix:grpo}

Group Relative Policy Optimization (GRPO)~\cite{shao2024deepseekmathpushinglimitsmathematical} is an adaptive reinforcement learning method that aligns policies via group-wise preference comparisons. Unlike DPO~\cite{rafailov2024directpreferenceoptimizationlanguage} and PPO~\cite{schulman2017proximalpolicyoptimizationalgorithms}, GRPO removes the need for a critic and standard reward models by using group-relative rewards from dynamic response clusters. This approach improves training stability, efficiency, and generalization, making it particularly suitable for multi-objective alignment tasks.

Although DPO also simplifies the reward modeling process from a modular perspective, it still faces certain limitations. Specifically, it relies on high-quality pairwise preference datasets and is sensitive to noisy preference data. In contrast, GRPO generates multiple responses within a single group and computes reward values based on intra-group comparisons. This allows for more natural and robust model training. 

During GRPO training, model-generated responses must be scored without relying on ground-truth outputs. However, the vanilla model lacks the ability to produce task-relevant responses, particularly in our setup, where the goal is to tag responses to malicious queries containing trigger words.

To address this, we perform a small number of epochs of supervised fine-tuning on the adapter layer using the loss function defined in~\autoref{eq:sft}. This warm-up phase ensures that the model can generate meaningful candidate responses before GRPO training begins. After this step, we proceed with GRPO training. We denote the objective function after this warm-up phase as follows:

\begin{equation}
\begin{aligned}
\theta^* &= \arg\max_{\theta_a} \;
\mathbb{E}_{\substack{x \sim D_* \\ G \sim \mathcal{G}(x)}} \Bigg[
\sum_{y_i, y_j \in G} \mathbb{I}_{y_i \succ y_j} \cdot 
\log \sigma \Bigg( 
\operatorname{clip}\Bigg(
\beta \cdot \bigg( \\
&\quad
\log \frac{\mathcal{M}_{\theta}(y_i \mid x)}{\mathcal{M}_{\theta^*}(y_i \mid x)} 
- 
\log \frac{\mathcal{M}_{\theta}(y_j \mid x)}{\mathcal{M}_{\theta^*}(y_j \mid x)}
\bigg),\;
-\delta,\; \delta
\Bigg)
\Bigg)
\Bigg] \\
&\quad - \gamma \cdot \mathbb{D}_{\mathrm{KL}} \left( 
\mathcal{M}_{\theta^*}(y \mid x) \;\middle\|\; \mathcal{M}_{\theta}(y \mid x) 
\right)
\end{aligned}
\end{equation}
where $\mathcal{G}$ represents the set of group response for all queries, $\mathbb{I}_{y_i \succ y_j}$ indicate reward function preference for $y_i$ over $y_j$, and $\gamma$ is a hyper-parameter that is used to control group-wise optimization with respect to KL regularization.

GRPO also constrains deviation from the reference model via a KL divergence regularization term in the loss function. This penalizes model shifts during training. We posit that this design can also contributes to improved stealthiness in the instrumented model.

\section{More Details about Detection Scenario} \label{appendix:Scenario}

In~\autoref{tab:diff_settings}, we discuss four trigger--tag settings. `ExT' refers to an explicit trigger, `ExG' to an explicit tag, `ImT' to an implicit trigger, and `ImG' to an implicit tag. We observe that using explicit triggers and tags typically leads to the highest detection performance.

\section{More Details about Jailbreak Testing} \label{appendix:jailbreak}

In this section, we simulate attacks by malicious users using $250$ jailbreak prompts to evaluate the robustness of \base, \core, and \pro. As shown in~\autoref{tab:jailbreak}, our method remains robust to jailbreak attempts under all three settings. The trigger--tag association is not compromised by these adversarial prompts.

\begin{table}[!h]
    \centering
    \caption{The detection accuracy after the instrumented model is exposed to jailbreak prompts.}
    \label{tab:jailbreak}
    \small
    \resizebox{0.45\textwidth}{!}{
    \begin{tabular}{c|cccc}
    \toprule
         \multirow{1}{*}{Strategy}&$32$ & $64$ & $128$ & $256$ \\ \midrule
         \base & $0.850$ & $0.825$ & $0.87$ & $0.86$ \\
         \core & $1.000$ & $1.000$ & $0.99$ & $0.995$ \\
         \pro & $0.820$ & $0.940$ & $0.945$ & $0.830$\\
    \bottomrule
    \end{tabular}
    }
\end{table}

\end{document}